\documentclass[a4paper,11pt]{article}
\pdfoutput=1
\usepackage{jheppub}

\title{On the two-body decay processes of the predicted three-body $K^*(4307)$ resonance}
\date{\today}
\author{Xiu-Lei Ren,}
\emailAdd{xiulei.ren@rub.de}
\affiliation{Ruhr-Universit\"{a}t Bochum, Fakult\"{a}t f\"{u}r Physik und Astronomie, Institut f\"{u}r Theoretische Physik II, D-44780 Bochum, Germany.}

\author{Brenda B. Malabarba,}
\emailAdd{brenda@if.usp.br}
\affiliation{Instituto de F\'isica, Universidade de S\~ao Paulo, C.P. 66318, 05389-970 S\~ao 
Paulo, S\~ao Paulo, Brazil.}

\author{K. P. Khemchandani,}
\emailAdd{kanchan.khemchandani@unifesp.br}
\affiliation{Universidade Federal de S\~ao Paulo, C.P. 01302-907, S\~ao Paulo, Brazil.}

\author{A. Mart\'inez Torres}
\emailAdd{amartine@if.usp.br}
\affiliation{Instituto de F\'isica, Universidade de S\~ao Paulo, C.P. 66318, 05389-970 S\~ao 
Paulo, S\~ao Paulo, Brazil.}

\abstract{
In a recent theoretical work, Phys. Lett. B785, 112 (2018), we proposed that a $K^*$ resonance with hidden charm content arises from the $KD\bar D^*$ dynamics, where the $D\bar D^*$ system is treated as a $Z_c(3900)$ or $X(3872)$. With the motivation of determining its further properties, which can be observed in experiments, we now present a calculation of the decay processes of this $K^*$, namely $K^*(4307)$, to two-body channels.
Particularly, we consider the decay channels $J/\psi K^*(892)$, $\bar{D}D_s$, $\bar{D}D_s^*$ and $\bar{D}^* D_s^*$. The mechanisms of the decay to these channels involve triangular loops and are a consequence of the internal structure of the state. Thus, the values found for the decay widths of the proposed $K^*(4307)$ are related to its nature and should be valuable for carrying on an experimental study of the $K^*(4307)$. A $K^*$ state with such a mass (in the charmonium region) and quantum numbers is a clear manifestation of an exotic meson, since, having hidden charm (i.e., a $c\bar c$ pair), its mass and quantum numbers can not be explained within a quark-antiquark description. }

\begin{document}

\maketitle
\section{Introduction}
The existence of exotic mesons and baryons, whose masses, widths and/or quantum numbers can not be explained within the constituent quark model of Gell-Mann and Zweig, is one of the peculiar characteristics of Quantum Chromodynamics which has been, and still is being, intensively explored in experiments and in theory. Typical examples are: the scalar nonet in the meson sector, which includes the $f_0(500)$, $\kappa(800)$, $f_0(980)$, $a_0(980)$ states~\cite{Jaffe:1976ig,Weinstein:1982gc,vanBeveren:1986ea,Tornqvist:1995kr,Oller:1997ng,Oller:1998hw}, and the $\Lambda(1405)$ in the baryon sector~\cite{Dalitz:1959dn,Dalitz:1960du,Kaiser:1995eg,Oset:1997it,Meissner:1999vr}. With the increase of the accessible energy range by the experimental facilities, claims for the observation of such states, especially in the heavy quark sector, with a hidden charm content,  started appearing in the last decade, as the so called $X$, $Y$ and $Z$ families (see, e.g., Refs.~\cite{Klempt:2007cp, Brambilla:2010cs,Hosaka:2016pey, Oset:2016lyh, Lebed:2016hpi,Chen:2016qju,Olsen:2017bmm,Guo:2017jvc,Liu:2019zoy} for reviews on the topic).  In case of the $Z$ family, consisting of charged particles with masses in the charmonium mass range, 3.9$-$4.2 GeV, at least two quarks and two antiquarks are necessarily required, with a $c\bar c$ pair being responsible for their heavier masses. The isoscalar partners, belonging to the $X$ and $Y$ families, are also categorized as exotic, not due to the fact that to obtain their quantum numbers we need to invoke a different structure to that of $q\bar q$, but because their masses and widths cannot be explained within the traditional constituent quark model~\cite{Chen:2016qju,Guo:2017jvc}.

All these heavy exotic mesons found experimentally in the recent years share a common feature: they are mesons with no strangeness. A glance at the Particle Data Book (PDB)~\cite{Tanabashi:2018oca} shows a low activity in the strange pseudoscalar and vector meson sectors since the last 30 years: in the pseudoscalar sector, the last $I(J^P)=1/2(0^-)$ Kaon state reported corresponds to $K(1830)$. Its existence was claimed in 1983 from a partial wave analyses of the $K^-\phi$ system produced in the reaction $K^- p\to K^+ K^- K^- p$~\cite{Armstrong:1982tw}, and, recently, the LHCb collaboration took it into account in the amplitude analysis of the $B^+\to J/\psi \phi K^+$ decay~\cite{Aaij:2016iza}. Similarly, in the vector sector, the latest $I(J^P)=1/2 (1^-)$ $K^*$  state listed in Ref.~\cite{Tanabashi:2018oca} is the $K^*(1680)$, whose existence dates to experiments and partial wave analysis performed during 1978-1988~\cite{Estabrooks:1977xe,Etkin:1980me,Aston:1986jb,Aston:1987ir}. As in the case of $K(1830)$ also, the LHCb collaboration has recently considered its existence in the analysis of the amplitude for the decay process $B^+\to J/\psi \phi K^+$~\cite{Aaij:2016iza}. And, overall, the final excited state in the meson sector, with nonzero strangeness quantum number, reported in the PDB corresponds to $K(3100)$, whose quantum numbers are unknown, and which was observed in several $\Lambda\bar p (\bar \Lambda p)+$pions reactions during the years 1986-1993~\cite{Aaij:2016iza}.

In view of such a panorama, it is worth to explore whether or not there could be another family member to be added to the already known $X$, $Y$, $Z$ families whose members will also have masses in the charmonium mass range, i.e., $\sim 3-4$ GeV, but nonzero strangeness. Such states  are manifestly exotic, since within a quark description, we will need at least a $c\bar c$ pair as well as a $s$ quark and a light antiquark ($\bar u$, $\bar d$) to account for their masses and quantum numbers. Surprisingly, although being currently accessible, the existence of such states has not been yet explored experimentally. But formation of such states has been claimed theoretically very recently using different models: in Ref.~\cite{Ma:2018vhp}, the $DD^* K$ system was studied by solving the Schr\"odinger equation and considering a pion exchange potential model to describe the interactions between the pairs forming the three-body system. As a result, a bound state with mass $4317.92^{+3.66}_{-4.32}$ MeV was obtained. Considering $G$-parity arguments, the authors of Ref.~\cite{Ma:2018vhp} claim also the existence of a $D\bar D^*K$ bound state with basically the same mass. In Ref.~\cite{Ren:2018pcd}, the $D\bar D^* K$ system was studied by solving the Faddeev equations under the fixed center approximation~\cite{Kamalov:2000iy,Xie:2010ig,Roca:2010tf,MartinezTorres:2010ax,Bayar:2011qj,Bayar:2015oea,Debastiani:2017vhv,Ren:2018qhr}. In this case,  the interaction between the particles in the two-body subsystems were obtained by solving the Bethe-Salpeter equation in coupled channels with a kernel determined from an effective field theory implementing symmetries like the chiral symmetry~\cite{Gasser:1983yg,Gasser:1984gg} or the heavy quark spin symmetry~\cite{Voloshin:1978hc,Isgur:1989vq,Burdman:1992gh}. Under such an approach, the states $D^*_{s0}(2317)$, $X(3872)$ and $Z_c(3900)$ are generated from the coupled channel dynamics and are mainly $DK$ bound states in isospin 0, $D\bar D^*$ states in isospin 0 and 1, respectively~\cite{Gamermann:2006nm,Guo:2006fu,Nieves:2012tt,Aceti:2014uea}. As a consequence of the dynamics involved, a theoretical evidence for an $I(J^P)=1/2(1^-)$ $K^*$ state with a mass of $(4307\pm 2)-i(9\pm 2)$ MeV was obtained when the $D\bar D^*$ system clusters as $X(3872)$ or $Z_c(3900)$.

Theoretically, the attraction in the $DK$ and $D\bar D^*$ subsystems, which leads to the generation of the $D^*_{s0}(2317)$, $X(3872)$ and $Z_c(3900)$ states, constitutes a compelling argument in favor of the existence of such exotic $K^*$ state with a mass around 4.3 GeV and hidden charm. Experimentally, observation of such $K^*$ state should be possible in the current  facilities and it would constitute an exciting  novelty in the Kaonic spectroscopy and in that of the exotic mesons.

In the present work, we continue with the investigation of the properties of the $K^*$ state predicted in Ref.~\cite{Ren:2018pcd} and calculate the decay widths to several open two-body channels. Particularly, we consider the channels $J/\psi K^*(892)$, $\bar D D^*_s$, $\bar D^* D^*_s$ and $\bar D D_s$, which are the most relevant ones, based on the nature of $K^*(4307)$. This information should be reliable for experimental searches of the $K^*$ state proposed in Ref.~\cite{Ren:2018pcd}, since the decay mechanism of the state is linked to the internal structure of the decaying particle.
\begin{figure}
\centering
\includegraphics[width=0.8\textwidth]{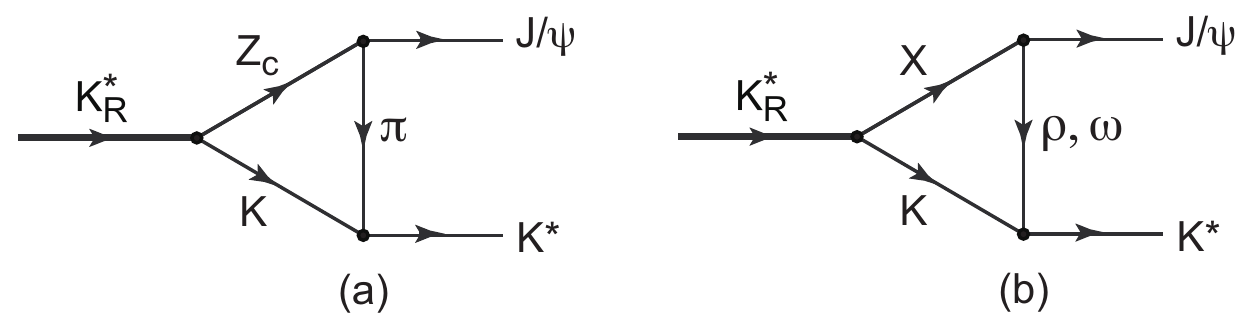}
\caption{Decay mechanisms of the $K^*_R$ state predicted in Ref~\cite{Ren:2018pcd} to the $J/\psi K^*$ channel. The vertex $X\to J/\psi\rho(\omega)$ on the diagram (b) involves yet another triangular loop, as shown in Fig.~\ref{Fig:XtoJpsi_vertex}}\label{Fig:compare}
\end{figure}

\section{Theoretical Framework}
\begin{figure}
\centering
\includegraphics[width=0.8\textwidth]{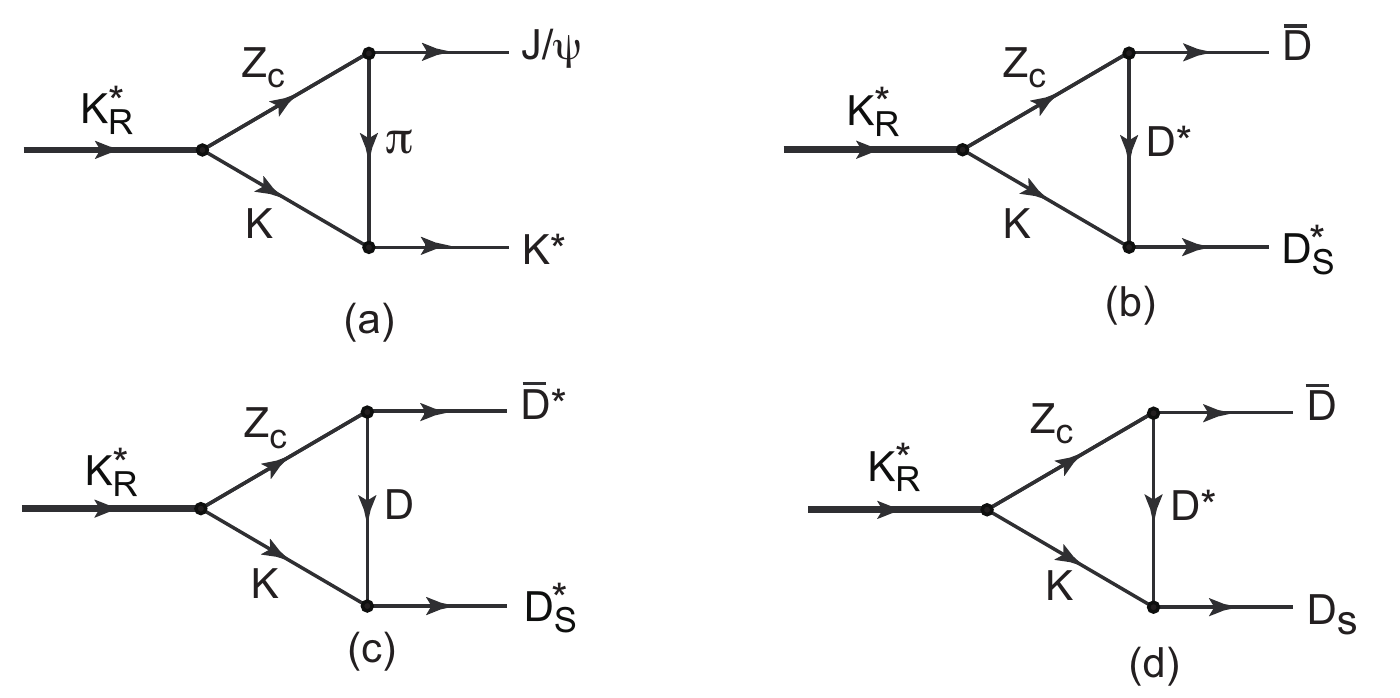}
\caption{Main two-body decay channels for the $K^*_R$ state found in the theoretical investigation of Ref~\cite{Ren:2018pcd}.}\label{Fig:decaydiags}
\end{figure}

The coupled channel calculation of Ref.~\cite{Ren:2018pcd} shows that the rescattering of a Kaon with the $D$ and $\bar D^*$, which cluster to form $X(3872)$ in isospin 0 and $Z_c(3900)$ in isospin 1, generates a $I(J^P)=1/2(1^-)$ $K^*$ state with a mass around 4307 MeV, which is below the $KD\bar D^*$ threshold, thus, it is a bound state. When considering the width of $Z_c(3900)$, which is around 28 MeV, a width close to 18 MeV is found for the $K^*(4307)$ state. A $K^*$ state with such an internal structure can naturally decay to three-body channels, like $J/\psi\pi K$, since the state itself is obtained as a consequence of the three-body dynamics involved in the $KD\bar D^*$ system. However, it can also decay to two-body channels. In this latter case, due to the nature found for $K^*(4307)$ in Ref.~\cite{Ren:2018pcd}, such a decay mechanism can proceed through triangular loops (see Fig.~\ref{Fig:compare}) and we can have as main decay channels $J/\psi K^*(892)$, $\bar D D^*_s$, $\bar D^* D^*_s$, and $\bar D D_s$ (see Fig.~\ref{Fig:decaydiags}). In order to avoid confusion between $K^*(4307)$ and $K^*(892)$ and to simplify the notation, we shall, henceforth, denote the former as $K^*_R$ and the latter as $K^*$. 

\begin{figure}
\centering
\includegraphics[width=0.4\textwidth]{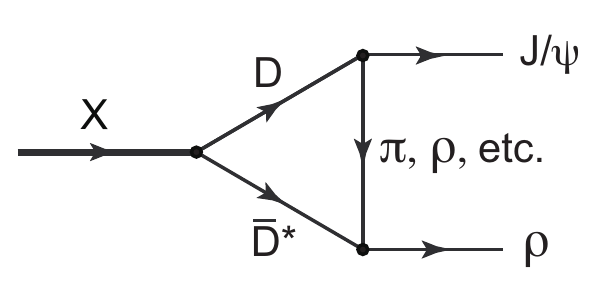}
\caption{Decay mechanism of $X$ to the $J/\psi\rho$ channel in an approach in which $X$ is obtained from the $D\bar D^*$ interaction~\cite{Aceti:2012cb}.}\label{Fig:XtoJpsi_vertex}
\end{figure}

\begin{figure}
\centering
\includegraphics[width=0.4\textwidth]{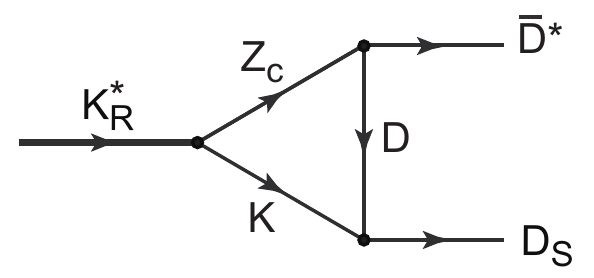}
\caption{Decay process of $K^*_R$ into the $\bar D^* D_s$ channel.}\label{Fig:triangle-of-Ds+barDx}
\end{figure}

From the results of Ref.~\cite{Ren:2018pcd}, the coupling of $K^*_R$ to  $KZ_c(3900)$ is around 4 times bigger than that to $KX(3872)$, thus, when calculating the decay width of $K^*_R$ (which is proportional to the squared coupling of $K^*_R$ to $KZ_c$ or $KX$), the contribution arising from the diagram shown in Fig.~\ref{Fig:compare}(b) is negligible when compared to the one coming from the diagram in Fig.~\ref{Fig:compare}(a). On top of that, for the decay process $K^*_R\to J/\psi K^*$, the vertex $X\to J/\psi \rho (\omega)$ shown in Fig.~\ref{Fig:compare}(b) involves yet another triangular loop~\cite{Aceti:2012cb} (see Fig.~\ref{Fig:XtoJpsi_vertex}) and such a vertex produces a contribution much smaller than that of the vertices $Z_c\to J/\psi\pi$, $\bar D D^*$, $\bar D^* D$, since $Z_c(3900)$ couples directly to $J/\psi\pi$, $\bar D D^*-\text{c.c}$ (where c.c means complex conjugate)~\cite{Aceti:2014uea}, at the tree level.
It is also interesting to notice that, with the internal structure found in Ref.~\cite{Ren:2018pcd} for the predicted $K^*_R$, the decay process $K^*_R\to \bar D^* D_s$ could also be contemplated, but it would involve a three pseudoscalar vertex (see Fig.~\ref{Fig:triangle-of-Ds+barDx}), resulting in a null amplitude. 
\subsection{Determination of the vertices}
\begin{figure}
\centering
\includegraphics[width=0.7\textwidth]{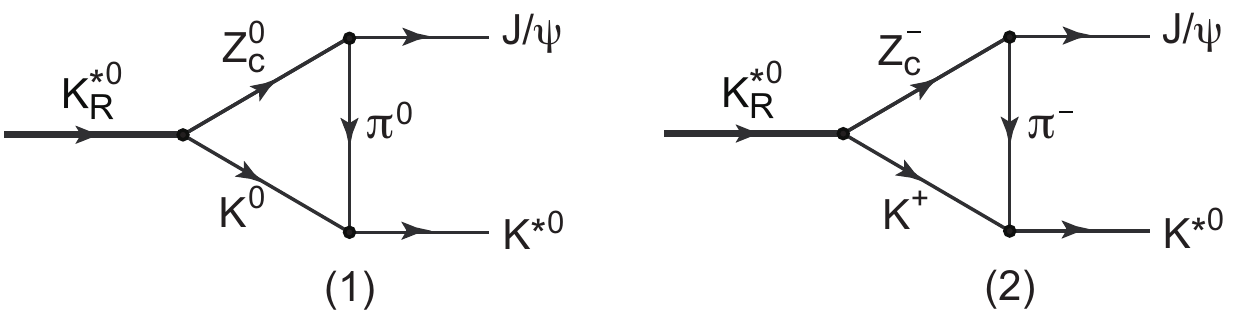}
\caption{Contributions related to the diagram (a) of Fig.~\ref{Fig:decaydiags} for the decay mechanism $K^{*0}_R\to J/\psi K^{*0}$.}\label{JpsiKstar}
\end{figure}

Let us then start evaluating the contribution arising from the diagrams shown in Fig.~\ref{Fig:decaydiags}. Considering the decay of a neutral $K^{*0}_R$ into a $J/\psi$ and a $\pi^0$, we have two diagrams contributing to each of the processes shown in Fig.~\ref{Fig:decaydiags}: in one of the diagrams, the primary vertex is $K^{*0}_R\to K^0 Z^0_c$ while in the other it is the vertex $K^{*0}\to K^+ Z^-_c$. We illustrate these two contributions in Fig.~\ref{JpsiKstar} for the decay process $K^{*0}_R\to J/\psi K^{*0}$.

\begin{figure}
\centering
\includegraphics[width=0.4\textwidth]{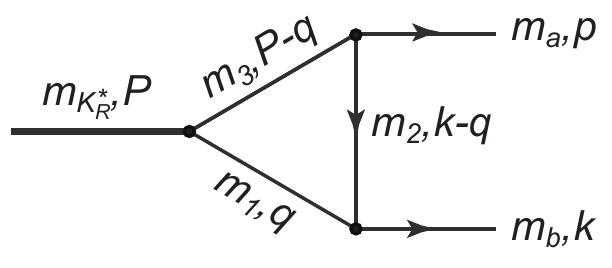}
\caption{Momenta and mass assignment in the decay process of the $K^*_R$ state.}\label{tri}
\end{figure}

To evaluate these diagrams, we need several vertices involving vector and pseudoscalar mesons. The contribution for the $K^*_R\to KZ_c$ vertices in Fig.~\ref{JpsiKstar} can be written in terms of the polarization vectors $\epsilon^\mu_{K^*_R} $ and $\epsilon^\mu_{Z_c}$ associated with the vector mesons $K^*_R$  and $Z_c$, respectively, and the coupling of $K^*_R$ to the $KZ_c(3900)$ channel as 
\begin{align}
t_{K_R^{*0}\to K^0Z^0_c} &= g_{K_R^{*0}\to K^0 Z^0_c}\, \epsilon_{K_R^{*0}}(P)\cdot \epsilon_{Z^0_c}(P-q),\nonumber\\
t_{K_R^{*0}\to K^+Z^-_c} &= g_{K_R^{*0}\to K^+ Z^-_c}\, \epsilon_{K_R^{*0}}(P)\cdot \epsilon_{Z^-_c}(P-q),\label{Eq:VKR}
\end{align}
where the four momenta and masses assigned to the particles are as shown in Fig.~\ref{tri}. 
The couplings $g_{K_R^{*0}\to K^0 Z^0_c}$ and $g_{K_R^{*0}\to K^+ Z^-_c}$ in Eq.~(\ref{Eq:VKR}) can be obtained from the isospin 1/2 scattering matrix, $T_{(KZ_c)_{\frac{1}{2}}}$, determined in Ref.~\cite{Ren:2018pcd}. To do this, we consider, a Breit-Wigner expression for this $T$-matrix in an energy region around the mass $m_{K^*_R}$ of the state, i.e.,
\begin{equation}\label{Eq:tTZcK}
T_{(KZ_c)_{\frac{1}{2}}} \simeq \frac{g_{K_R^*\to (KZ_c)_{\frac{1}{2}}}^2}{s-M_{K_R^*}^2+i M_{K_R^*}\Gamma_{K_R^*}} \vec{\epsilon}_{Z_c}\cdot\vec{\epsilon}_{Z_c}\equiv \tilde{T}_{Z_c K}\vec{\epsilon}_{Z_c}\cdot\vec{\epsilon}_{Z_c},
\end{equation}
and we can get $g_{K_R^*\to (KZ_c)_{\frac{1}{2}}}$, i.e., the coupling of $K^*_R$ to  the isospin 1/2 state $K Z_c$, from the residue of $T_{(KZc)_{\frac{1}{2}}}$ at the pole position in the complex energy plane. Alternatively,  since the decay width of $K^*_R$ is proportional to $|g_{K_R^*\to (KZ_c)_{\frac{1}{2}}}|^2$, we can estimate such a value directly from Eq.~(\ref{Eq:tTZcK}), by considering the limit $s\to M^2_{K^*_R}$~\cite{Nagahiro:2008um}
\begin{equation}
|g_{K_R^*\to (KZ_c)_{\frac{1}{2}}}|=\Big|\sqrt{iM_{K_R^*}\Gamma_{K_R^*} \tilde{T}_{(KZ_c)_{\frac{1}{2}}}}\Big|\simeq22143~\text{MeV}.
\end{equation}
Once we have the value of $g_{K_R^*\to (KZ_c)_{\frac{1}{2}}}$, the couplings  $g_{K_R^{*0}\to K^0 Z^0_c}$ and $g_{K_R^{*0}\to K^+ Z^-_c}$ can be related to $g_{K_R^*\to (KZ_c)_{\frac{1}{2}}}$ by using the fact that
\begin{align}
|KZ_c; I=\frac{1}{2},I_3=-\frac{1}{2}\rangle=-\frac{1}{\sqrt{3}}|K^0 Z^0_c\rangle+\sqrt{\frac{2}{3}}|K^+ Z^-_c\rangle,\label{KZc}
\end{align}
where we use the phase convention $|K^-\rangle=-|\frac{1}{2},-\frac{1}{2}\rangle$. In this way, from Eq.~(\ref{KZc}),
\begin{align}
g_{K^{*0}_R\to K^0 Z^0_c}&=-\frac{1}{\sqrt{3}}g_{K^*_R\to (K Z_c)_{\frac{1}{2}}},\nonumber\\
g_{K^{*0}_R\to K^+ Z^-_c}&=\sqrt{\frac{2}{3}}g_{K^*_R\to (K Z_c)_{\frac{1}{2}}}.\label{gcharge}
\end{align}
Using isospin average masses for the particles belonging to the same isospin multiplet and Eq.~(\ref{gcharge}), we can write Eq.~(\ref{Eq:VKR}) as
\begin{align}
t_{K_R^*\to KZ_c} &= C_{KZ_c}\, g_{K_R^*\to (KZ_c)_{\frac{1}{2}}}\, \epsilon_{K^*_R}(P)\cdot \epsilon_{Z_c}(P-q),
\end{align}
with 
\begin{align}
C_{KZ_c}=\left\{\begin{array}{c}-1/\sqrt{3}\quad \text{for $K^{*0}_R\to K^0 Z^0_c$},\\ 
\sqrt{2/3}\quad\text{for $K^{*0}_R\to K^+ Z^-_c$}.\end{array}\right.
\end{align}

Next, we need the vertices $Z_c\to J/\psi \, \pi$, $\bar D D^*$, $\bar D^* D$ for different charge combinations. As shown in Ref.~\cite{Aceti:2014uea}, a state with mass around 3872 MeV and 30 MeV of width is generated from the dynamics present in the $D\bar D^*+\text{c.c.}$ (c.c. means complex conjugate) and $J/\psi\pi$ coupled channel system in isospin 1 and positive $G$-parity. This state can be related to $Z_c(3900)$~\cite{Aceti:2014uea}. 

A comment regarding this latter state is here in order: the nature of $Z_c(3900)$ is still under debate. Experimental investigations seem to report two states with $J^P=1^+$ around 3900 MeV, $Z_c(3900)$~\cite{Ablikim:2013mio,Liu:2013dau,Xiao:2013iha} and $Z_c(3885)$~\cite{Ablikim:2013xfr,Ablikim:2015swa}. It is still not clear if these states are two different ones or are the same. The lattice investigations~\cite{Prelovsek:2013xba,Prelovsek:2014swa,Chen:2014afa}, on the other hand, do not seem to find an evidence for the existence of a molecular state around 3900 MeV. However, the analysis made in Ref.~\cite{Albaladejo:2016jsg} shows that the lattice data is compatible with the existence of the $Z_c(3900)$ resonance. Further, the latest experimental investigations continue to find signals of a state with mass near 3900 MeV in different processes, such as $B$-decays~\cite{Abazov:2018cyu}, $\eta_c\rho$ invariant mass spectra~\cite{Yuan:2018inv}, etc. In spite of the debate, both experimental and theoretical investigations indicate that the $D\bar D^*$ interaction in isospin 1, spin-parity $1^+$ is attractive in nature and produces a peak in the cross sections of the relevant processes. In our study, the $Z_c(3900)$ we refer to is the state arising from the $D\bar D^*$ and coupled channel dynamics found in Ref.~\cite{Aceti:2014uea}.

Following the approach of Ref.~\cite{Aceti:2014uea}, we can write
\begin{align}
t_{Z_c\to J/\psi\, \pi} &= C_{J/\psi\pi} g_{Z_c\to (J/\psi\, \pi)_{1}} \, \epsilon_{Z_c}(P-q)\cdot\epsilon_{J/\psi}(p),\nonumber\\
t_{Z_c\to \bar D\, D^*}&= C_{\bar D\, D^*}g_{Z_c\to(\bar D D^*)_{1}} \, \epsilon_{Z_c}(P-q)\cdot \epsilon_{D^*}(p),\label{Eq:VZcJ}\\
t_{Z_c\to \bar D^*D}&= C_{\bar D^* D}g_{Z_c\to(\bar D D^*)_{1}} \, \epsilon_{Z_c}(P-q)\cdot\epsilon_{\bar D^*}(p),\nonumber
\end{align}
where we have defined 
\begin{align}
g_{Z_c\to(\bar D D^*)_{1}}=\frac{1}{\sqrt{2}}g_{Z_c\to \frac{1}{\sqrt{2}}[(\bar D D^*)_{1}+\text{c.c}]}.
\end{align}
The subscript 1 in the above equation indicates the total isospin of the $\bar D D^*$ system. The $C_{\bar D\, D^*}$ and $C_{\bar D^* D}$ coefficients in Eq.~(\ref{Eq:VZcJ}), which relate the $Z_c$ state to the $\bar D\, D^*$ and $D\, \bar D^*$ states in the charge basis, are given by
\begin{align}
C_{\bar D\, D^* (\bar D^* D)}=\left\{\begin{array}{c}1,\quad\text{for $Z^+_c\to \bar D^0 D^{*+}\,(\bar D^{*0}D^+)$},\\
1/\sqrt{2},\quad\text{for $Z^0_c\to D^-D^{*+}\,(D^{*-}D^+)$},\\
-1/\sqrt{2},\quad\text{for $Z^0_c\to \bar D^0 D^{*0}\,(\bar D^{*0}D^0)$},\\
-1,\quad\text{for $Z^-_c\to D^- D^{*0}\,(D^{*-}D^0 )$},\end{array}\right.
\end{align}
where we have used the isospin phase convention $|D^{*0}\rangle=-|\frac{1}{2},-\frac{1}{2}\rangle$ and $|D^0\rangle=-|\frac{1}{2},-\frac{1}{2}\rangle$. In case of pions, we follow the isospin phase convention $|\pi^+\rangle=-|1,1\rangle$. In this way, $C_{J/\psi\pi}=1$ for the processes $Z^0_c\to J/\psi \pi^0$ and $Z^-_c\to J/\psi \pi^-$.

The couplings in Eq.~(\ref{Eq:VZcJ}) can be obtained from the residue of the isospin 1 two-body scattering matrix determined in Ref.~\cite{Aceti:2014uea} in the complex energy plane. We have calculated them and obtain
\begin{align}
|g_{Z_c\to (J/\psi\pi)_{1}}|\simeq 3715~\text{MeV},\quad |g_{Z_c\to \frac{1}{\sqrt{2}}[(\bar D \, D^*)_{1}+\text{c.c.}]}|\simeq8149~\text{MeV}.
\end{align}

Other vertices needed to evaluate the contribution of the diagrams shown in Fig.~\ref{Fig:decaydiags} are $K\,\pi \to K^*$, $D^* K\to D_s$ and $DK\to D^*_s$. To determine these contributions, we use the effective Lagrangian $\mathcal{L}_{PPV}$~\cite{Bando:1987br,Oset:2009vf} involving two pseudoscalars and a vector meson
\begin{equation}\label{Eq:LagPPV}
	\mathcal{L}_{PPV} = -ig\langle V^\mu \, [P,\partial_\mu P]\rangle,
\end{equation} 
with $V^\mu$ and $P$ being matrices containing the corresponding vectors and pseudoscalar fields, 
\begin{align}
 V_\mu &= 
\begin{pmatrix}
\frac{\omega + \rho^0}{\sqrt{2}} & \rho^+ & K^{*+} & \bar{D}^{*0} \\
\rho^- & \frac{\omega -\rho^0}{\sqrt{2}} & K^{*0} & D^{*-} \\
K^{*-} & \bar{K}^{*0} & \phi & D_s^{*-} \\
D^{*0} & D^{*+} & D_s^{*+} & J/\psi 
\end{pmatrix}_\mu,\nonumber\\
P &= 
\begin{pmatrix}
\frac{\eta}{\sqrt{3}} + \frac{\eta'}{\sqrt{6}} + \frac{\pi^0}{\sqrt{2}} & \pi^+ & K^+ & \bar{D}^0 \\
\pi^- & \frac{\eta}{\sqrt{3}} + \frac{\eta'}{\sqrt{6}} - \frac{\pi^0}{\sqrt{2}} & K^0 & D^- \\
K^- & \bar{K}^0 & -\frac{\eta}{\sqrt{3}} + \sqrt{\frac{2}{3}} \eta' & D_s^- \\
D^0 & D^+ & D_s^+ & \eta_c	
\end{pmatrix},
\end{align}
respectively. The coupling $g$ in Eq.~(\ref{Eq:LagPPV}) is given by $m_V/(2f_\pi)\simeq 4.41$, with $m_V\simeq 815$ MeV being an average mass for the vector mesons $\rho$, $\omega$ and $K^*$ and $f_\pi=92.4$ MeV being the pion decay constant. While this value of the coupling produces a theoretical width of the $K^{*+}$ meson, which comes basically from the decay processes $K^{*+}\to K^0\pi^+$, $K^+\pi^0$, compatible with the experimental result, it underestimates the width of the $D^{*+}$ meson, obtained from the processes $D^{*+}\to D^0\pi^+$, $D^+\pi^0$. In this latter case, as shown in Ref.~\cite{Aceti:2014kja}, arguments based on the heavy quark symmetry establish that $g\to m_{D^*} g/m_{K^*}\simeq 9.9$ when using the Lagrangian in Eq.~(\ref{Eq:LagPPV}) for describing processes involving heavy pseudoscalar and vector mesons. Having this in mind and using Eq.~(\ref{Eq:LagPPV}), we get the following amplitudes for the above mentioned vertices
\begin{align}
t_{K\pi \to K^*} &= -2\, C_{K\pi}\,g_L \, q\cdot\epsilon_{K^*}(k),\nonumber\\
t_{KD\to D_s^*} &= -2\, C_{KD}\,g_H\, q\cdot \epsilon_{D_s^*}(k),\label{Eq:VKX}\\
t_{KD^*\to D_s} &= C_{KD^*}\,g_H\, (k+q)\cdot \epsilon_{D_s^*}(k-q).\nonumber
\end{align}
In the above equations, $g_L=4.41$ and $g_H=9.9$, and the coefficients $C_{K\pi}$, $C_{KD}$ and $C_{KD^*}$ are given by
\begin{align}
C_{K\pi}&=\left\{\begin{array}{c}1/\sqrt{2},\quad\text{for $K^0\pi^0\to K^{*0}$},\\
-1,\quad\text{for $K^+\pi^-\to K^{*0}$},\end{array}\right.
\end{align}
\begin{align}
C_{KD\,(KD^*)}=\left\{\begin{array}{c}-1,\quad\text{for $K^+ D^{0}\to D^{*+}_s$ ($K^+ D^{*0}\to D^+_s$)},\\
-1,\quad\text{for $K^0 D^{+}\to D^{*+}_s$ ($K^0 D^{*+}\to D^+_s$)}.\end{array}\right.
\end{align}

The last vertex whose contribution needs to be determined corresponds to $D^*K\to D^*_s$. To do this, we consider the effective Lagrangian $\mathcal{L}_{VVP}$~\cite{Bando:1987br,Meissner:1987ge} involving two vectors and a pseudoscalar meson
\begin{equation}
	\mathcal{L}_{VVP} = \frac{G'}{\sqrt{2}} \varepsilon^{\mu\nu\alpha\beta} \langle \partial_{\mu}V_{\nu}\partial_{\alpha} V_\beta P\rangle, \label{VVP}
\end{equation}
where the coupling $G'$ is  given by $\frac{3 {g'}^2}{4\pi^2 f_\pi}$ with $g'=-\frac{M_\rho}{2f_\pi} = -4.14$. Using Eq.~(\ref{VVP}), we can write
\begin{equation}
	t_{KD^*\to D_s^*} = C_{KD^*}G^\prime\,\varepsilon^{\mu\nu\alpha\beta}\, (k-q)_\mu \,k_\alpha \, \epsilon_{D^*,\nu}(k-q)\, \epsilon_{D_s^*,\beta}(k),\label{vDs}
\end{equation}
with  
\begin{align}
C_{KD^*}=\left\{\begin{array}{c}-\frac{1}{\sqrt{2}},\quad\text{for $K^+ D^{*0}\to D^{*+}_s$},\\
-\frac{1}{\sqrt{2}},\quad\text{for $K^0 D^{*+}\to D^{*+}_s$}.\end{array}\right.
\end{align}

\subsection{Triangular loops}
Once we have all the vertices associated with the decay mechanisms of $K^*_R$, we can evaluate the contributions related to the diagrams in Fig.~\ref{Fig:decaydiags}. We start with the process shown in Fig.~\ref{Fig:decaydiags}(a) and the two Feynman diagrams shown in Fig.~\ref{JpsiKstar} (for the $K^{*0}_R$ decay). Using the vertices given in Eqs.~(\ref{Eq:VKR}), (\ref{Eq:VZcJ}), (\ref{Eq:VKX}), the corresponding amplitude can be written as, 
\begin{align}\label{Eq:ta}
	t_{a} &=t^{(1)}_a+t^{(2)}_a= i\sqrt{6}\,g_{K_R^*\to (KZ_c)_{\frac{1}{2}}}\, g_{Z_c\to (J/\psi\,\pi)_{1}}\, g_L \, 
	\epsilon_{K_R^*}^\mu(P)\, \epsilon_{J/\psi}^\nu(p)\,\epsilon_{K^*}^\alpha(k)\nonumber\\
	&\quad\times \int\frac{d^4 q}{(2\pi)^4} 
	\frac{\left( -g^{\mu\nu} + \frac{(P-q)^\mu (P-q)^\nu}{m_{Z_c}^2}\right)q^\alpha}{ (q^2-m_K^2+i\epsilon) [(k-q)^2-m_\pi^2 + i\epsilon] [(P-q)^2-m_{Z_c}^2 + i\epsilon] } \nonumber\\
 &= i\sqrt{6}\,g_{K_R^*\to (KZ_c)_{\frac{1}{2}}}\, g_{Z_c\to (J/\psi\,\pi)_{1}}\, g_L \, 
	\epsilon_{K_R^*}^\mu(P)\, \epsilon_{J/\psi}^\nu(p)\,\epsilon_{K^*}^\alpha(k) \nonumber\\
	&\quad\times \left[-g_{\mu\nu}\, I_{\alpha}^1 - \frac{P_\nu}{m_{Z_c}^2}\, I_{\mu\nu}^2
	+\frac{1}{m_{Z_c}^2} \, I_{\mu\nu\alpha}^3  \right],
\end{align}
where we have introduced the three tensor integrals $I_\alpha^1,\, I_{\mu\nu}^2,\, I_{\mu\nu\alpha}^3$, which are defined as  
\begin{equation}\label{Eq:tensorInt}
I_{\mu_1 \mu_2,...,\mu_N}^N= 
\int\frac{d^4 q}{(2\pi)^4} 
	\frac{q_{\mu_1} q_{\mu_2} \cdots q_{\mu_N}}{(q^2-m_1^2+i\epsilon) [(k-q)^2-m_2^2 + i\epsilon] [(P-q)^2-m_3^2 + i\epsilon]}, 
\end{equation}
with $m_1$, $m_2$ and $m_3$ being the masses of the particles in the triangular loops shown in Fig.~\ref{Fig:decaydiags} (see Fig.~\ref{tri} for the corresponding four momenta labels).

Based on the Lorentz covariance, Eq.~(\ref{Eq:tensorInt}) can be written in terms of the external momentum $P$ and $k$. In particular, we have
\begin{align}
	I_\alpha^1 &= a_1^1\, P_\alpha + a_2^1\, k_\alpha,\nonumber\\
	I_{\mu\alpha}^2 &= a_1^2\, g_{\mu\alpha} + 
	a_2^2 \, P_\mu \, P_\alpha + a_3^2\, (P_\mu\, k_{\alpha} + k_\mu \, P_{\alpha} ) + a_4^2\, k_\mu\,k_\alpha, \label{Is}\\
   I_{\mu\nu\alpha}^3 &= a_1^3(g_{\mu\nu}P_{\alpha} + g_{\mu\alpha}P_\nu + g_{\nu\alpha}P_\mu) + a_2^3(g_{\mu\nu}k_\alpha + g_{\mu\alpha}k_\nu + g_{\nu\alpha}k_{\mu}) \nonumber\\
   &\quad+ a_3^3 P_\mu P_\nu P_\alpha + a_4^3(P_\mu P_\nu k_\alpha + P_\mu k_\nu P_\alpha + k_\mu P_\nu P_\alpha) \nonumber\\
   &\quad+ a_{5}^3 k_\mu k_\nu k_{\alpha} + a_{6}^3 (k_\mu k_\nu P_\alpha + k_\mu P_\nu k_\alpha + P_\mu k_\nu k_\alpha),\nonumber
\end{align}
which correspond to the standard Passarino-Veltman decomposition for tensor integrals~\cite{Passarino:1978jh}. The coefficients $a^i_{j}$ are scalars to be determined. Considering the Lorenz gauge and using $P=p+k$, the amplitude $t_a$ in Eq.~(\ref{Eq:ta}) can be further simplified to
\begin{align}\label{Eq:tasim}
	t_a &= i\sqrt{6}\,{m_{Z_c}^2}\,g_{K_R^*\to (KZ_c)_{\frac{1}{2}}}\, g_{Z_c\to (J/\psi\,\pi)_{1}}\, g_L \left[ (-m_{Z_c}^2 a_1^1 + a_1^3) \epsilon_{K_R^*}(P)\cdot \epsilon_{J/\psi}(p) \, p\cdot \epsilon_{K^*}(k) \right.\nonumber\\
	 &\quad+ (-a_1^2 + a_1^3 + a_2^3) \epsilon_{K_R^*}(P)\cdot\epsilon_{K^*}(k) \, 
	 k\cdot \epsilon_{J/\psi}(p) 
	 + a_2^3 k\cdot \epsilon_{K_R^*}(P) \, \epsilon_{J/\psi}(p)\cdot \epsilon_{K^*}(k) \nonumber\\
	 &\quad\left. + (-a_3^2 + a_4^3 + a_{6}^3) k\cdot \epsilon_{K_R^*}(P) \, k\cdot\epsilon_{J/\psi}(p) \, p\cdot\epsilon_{K^*}(k) \right], 
\end{align}
 and we need to determine seven coefficients, $a_{1}^1$, $a_{1}^2$, $a_{3}^2$, $a_1^3$, $a_2^3$, $a_{4}^3$, and $a_{6}^3$. To do this, the way of proceeding is:
first, by using Eq.~(\ref{Is}), we can contract the expressions in Eq.~(\ref{Is}) with the different Lorentz structures present there and get a system of coupled equations which can be solved. For example, from the expression of $I^1_\alpha$ in Eq.~(\ref{Is}), we have
 \begin{align}
 P\cdot I^1&=a_1^1P^2+a_2^1 P\cdot k,\nonumber\\
 k\cdot I^1&=a_1^1 k\cdot P+a_2^1 k^2.
 \end{align}
By solving this system of coupled equations, we can write $a^1_1$  as
\begin{align}
a_{1}^1&= \frac{k^2(\mathbb{PI}^1)-(k\cdot P)(\mathbb{KI}^1)}{k^2 P^2 - (k\cdot P)^2},\label{a11}
\end{align}
where 
\begin{align}
&\mathbb{PI}^1\equiv P^\mu I^1_\mu,\quad \mathbb{KI}^1\equiv k^\mu I^1_\mu,\quad \mathbb{GI}^2\equiv g^{\mu\alpha}I^2_{\mu\alpha}.\label{Is2}
\end{align}
Equation~(\ref{a11}) clearly shows that the $a^i_j$ coefficients depend on the mass of the decaying particle, $m_{K^*_R}$, the masses of the particles in the loops, $m_1$, $m_2$ and $m_3$, and the masses $m_a$ and $m_b$ of the particles to which $K^*_R$ can decay (see Fig.~\ref{tri}). For all the diagrams shown in Fig.~\ref{Fig:decaydiags}, $m_1=m_K$ and $m_3=m_{Z_c}$, and for the particular case of the diagram in Fig.~\ref{Fig:decaydiags}(a), $m_2=m_\pi$, $m_a=m_{J/\psi}$ and $m_b=m_{K^*}$. The next step consists in calculating the Lorentz scalar terms appearing in Eq.~(\ref{Is2}) directly from the definition in Eq.~(\ref{Eq:tensorInt}). For example, using Eq.~(\ref{Eq:tensorInt}), $\mathbb{PI}^1$ is given by 
\begin{align}
\mathbb{PI}^1= \int\frac{d q^0}{2\pi}\int \frac{d^3 q} {(2\pi)^3} \frac{P^0 q^0} {({q^0}^2-\omega_1^2+i\epsilon) [(k^0-q^0)^2-\omega_2^2 + i\epsilon] [(P^0-q^0)^2-\omega_3^2 + i\epsilon]}, \label{intPI1}
\end{align}
with 
\begin{align}
\omega_1=\sqrt{\vec{q}^{\,\,2}+m^2_1},\quad\omega_2=\sqrt{(\vec{k}-\vec{q})^{\,2}+m^2_2},\quad\omega_3=\sqrt{\vec{q}^{\,\,2}+m^2_3},
\end{align}
where we have used the rest frame of the decaying particle, for which $P^\mu=(P^0,\vec{0})=(m_{K^*_R},\vec{0})$ and
\begin{align}
k^0=\frac{{P^0}^2-m^2_a+m^2_b}{2P^0},\quad|\vec{k}|=\frac{\lambda^{1/2}({P^0}^2,m^2_a,m^2_b)}{2P^0}.
\end{align}
Next, we can use Cauchy's theorem to determine the $q^0$ integration of Eq.~(\ref{intPI1}), and we get

\begin{align}
	\mathbb{PI}^1&=  -i \int \frac{d^3 q}{(2\pi)^3}\,  
	P^0 \omega_1 \Big\{-{k^0}^2 P^0 \omega_2 + k^0 \omega_3 [(\omega_1 + \omega_3)(\omega_1 + 2\omega_2 + \omega_3) - {P^0}^2] \nonumber\\
	&\quad+ P^0 \omega_2 (\omega_1 + \omega_2) (\omega_1 + \omega_2 + 2\omega_3)\Big\} \, \frac{1}{2\,\omega_1\, \omega_2\, \omega_3\,(k^0 + \omega_1 + \omega_2 )}\nonumber\\
	&\quad \times \frac{1}{P^0+\omega_1+\omega_3} \frac{1}{\omega_1 -k^0 + \omega_2 - i\epsilon} \frac{1}{\omega_1-P^0+\omega_3-i\epsilon} \nonumber\\
	&\quad \times \frac{1}{k^0-P^0+\omega_2 +\omega_3 - i\epsilon} \frac{1}{P^0-k^0+\omega_3 + \omega_2 - i\epsilon}.\label{int}
\end{align}

Similarly, we can continue with the evaluation of the other $a^i_j$ coefficients of Eq.~(\ref{Eq:tasim}). The results are given in the appendix~\ref{ap}. Note that some of these $a^i_j$ coefficients, after performing the integration on the $q^0$ variable, involve integrals in $d^3q$ which are divergent. In such a case, we regularize the corresponding integral by introducing a cutoff  $\Lambda= 700$ MeV, which corresponds to the value used in Ref.~\cite{Ren:2018pcd} to get the resonance $K_R^*$ from the three-body $KD\bar{D}^*$ system. It is also interesting to notice that for the cases in which the $d^3q$ integration does not involve divergences, the upper limit for such integration is also naturally provided~\cite{Aceti:2015zva}, in this case, by the value of the cut-off used when regularizing the two-body loops involved in the generation of the $Z_c$ state from the interaction of $D\bar D^*$ and coupled channels, and which is also $\sim$700 MeV~\cite{Aceti:2014uea}.

\begin{figure}
\centering
\includegraphics[width=0.7\textwidth]{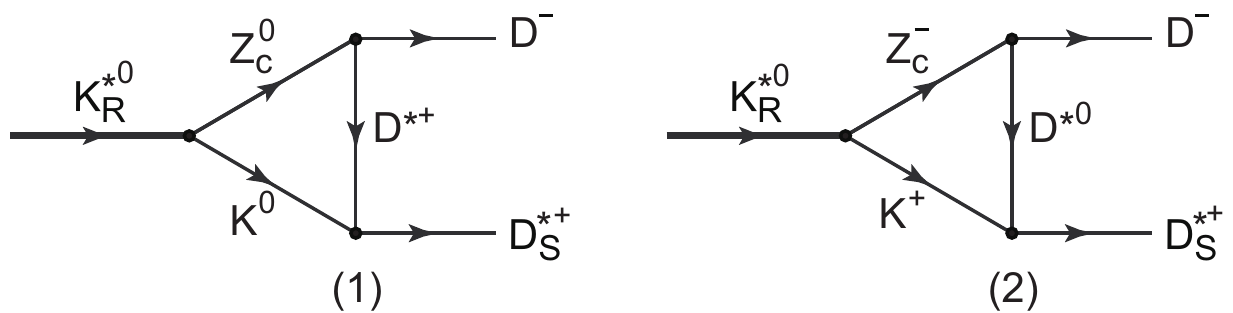}
\caption{Contributions associated with the diagram (b) of Fig.~\ref{Fig:decaydiags} for the decay mechanism $K^{*0}_R\to D^- D^{*+}_s$.}\label{DbarDstars}
\end{figure}

Let us consider now the decay mechanism shown in Fig.~\ref{Fig:decaydiags}(b) and the two Feynman diagrams contributing to it, which are shown in Fig.~\ref{DbarDstars}. In this case, considering Eqs.~(\ref{Eq:VKR}),~(\ref{Eq:VZcJ}) and (\ref{vDs}), the amplitude describing the process is given by
\begin{align}
	t_b &= t^{(1)}_b+t^{(2)}_b=i\,\frac{\sqrt{3}}{2} g_{K_R^*\to (KZ_c)_{\frac{1}{2}} }\, g_{Z_c\to(\bar{D}D^*)_{1}}\, G^\prime\, \epsilon_{K_R^*}^\mu(P)\,\epsilon_{D_s^*,\beta}(k)\nonumber\\
	&\quad\times\int\frac{d^4 q}{(2\pi)^4} \left( -g^{\mu\nu} + \frac{(P-q)^\mu (P-q)^\nu}{m_{Z_c}^2} \right) \left(-g_{\nu\lambda} + \frac{(k-q)_\nu (k-q)_\lambda}{m_{D^*}^2} \right) \nonumber\\
	&\quad\times \frac{\varepsilon^{\sigma\lambda\alpha\beta}\,(k-q)_\sigma \,k_\alpha}{(q^2-m_K^2+i\epsilon)[(k-q)^2-m_{D^*}^2+i\epsilon] [(P-q)^2-m_{Z_c}^2+i\epsilon]}\nonumber\\
&= i\,\frac{\sqrt{3}}{2} g_{K_R^*\to (KZ_c)_{\frac{1}{2}} }\, g_{Z_c\to(\bar{D}D^*)_{1}}\, G^\prime\, \epsilon_{K_R^*}^\mu(P)\,\epsilon_{D_s^*,\beta}(k)\, \varepsilon^{\sigma \lambda \alpha \beta} \left[ g_{\mu \lambda}k_\alpha I_\sigma^1+\frac{1}{M_{Zc}^2} P_\lambda k_\alpha I_{\mu \sigma}^2 \right],\label{tb}
\end{align}
where the Lorenz gauge and the antisymmetric properties of the Levi-Civita tensor have been used to get the last line. Using the decomposition in Eq.~(\ref{Is}) and considering once again the antisymmetric properties of the Levi-Civita tensor, Eq.~(\ref{tb}) can be written as
\begin{align}
t_b=i\,\frac{\sqrt{3}}{2} g_{K_R^*\to (KZ_c)_{\frac{1}{2}} }\, g_{Z_c\to(\bar{D}D^*)_{1}}\, G^\prime\, \epsilon_{K_R^*}^\mu(P)\,\epsilon_{D_s^*,\beta}(k)\, \varepsilon^{\sigma \lambda \alpha\beta} \left[g_{\mu \lambda} P_\sigma a^1_1+ \frac{1}{M_{Zc}^2} P_\lambda g_{\mu\sigma}a^2_1\right]k_\alpha,\label{Eq:tb}
\end{align}
where the coefficient $a^1_1$ can be obtained from Eq.~(\ref{a11}), where now, from Fig.~\ref{Fig:decaydiags}(b), $m_3=m_{D^*}$, $m_a=m_{\bar D}$, $m_b=m_{D^*_s}$, and the expression for $a^2_1$ can be found in the appendix~\ref{ap}.

\begin{figure}
\centering
\includegraphics[width=0.7\textwidth]{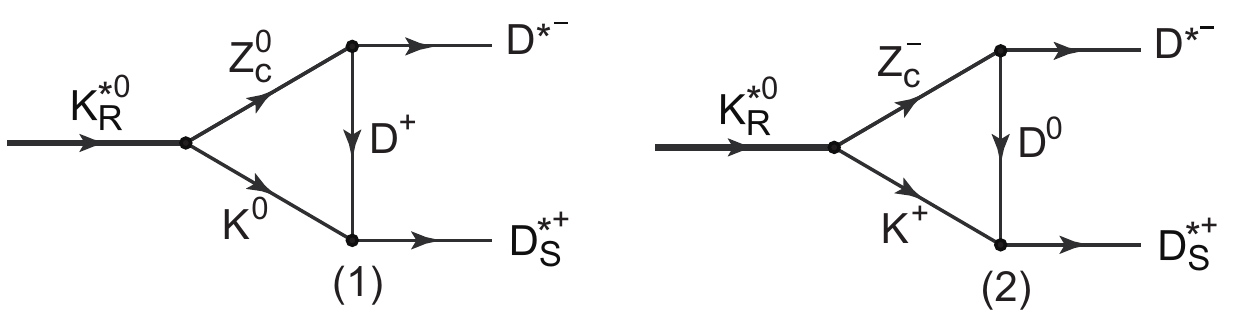}
\caption{Contributions associated with the diagram (c) of Fig.~\ref{Fig:decaydiags} for the decay mechanism $K^{*0}_R\to D^{*-} D^{*+}_s$.}\label{DbarstarDstars}
\end{figure}

Next, we continue with the evaluation of the process depicted in Fig.~\ref{Fig:decaydiags}(c). In this case, considering the diagrams shown in Fig.~\ref{DbarstarDstars} and using the results in Eqs.~(\ref{Eq:VKR}),~(\ref{Eq:VZcJ}),~(\ref{Eq:VKX}), the amplitude associated with such decay mechanism reads as
\begin{align}
  t_c &=t^{(1)}_c+t^{(2)}_c= -i\,\sqrt{6}\, g_{K_R^*\to (KZ_c)_{\frac{1}{2}}}\, g_{Z_c\to(\bar{D}D^*)_{1}}\, g_H\, \epsilon_{K_R^*}^\mu(P) \, \epsilon_{\bar{D}^*}^\mu(P-k)\, \epsilon_{D_s^*}^{\alpha}(k) \nonumber\\
  &\quad\times \int\frac{d^4 q}{(2\pi)^4} \frac{\left(-g_{\mu\nu} + \frac{(P-q)_\mu\,(P-q)_\nu}{m_{Z_c}^2}\right)\, q_\alpha}{(q^2-m_K^2+i\epsilon)[(k-q)^2-m_D^2+i\epsilon][(P-q)^2-m_{Z_c}^2+i\epsilon]}\nonumber\\
  &=  -i\,\sqrt{6}\, g_{K_R^*\to (KZ_c)_{\frac{1}{2}}}\, g_{Z_c\to(\bar{D}D^*)_{1}}\, g_H\, \epsilon_{K_R^*}^\mu(P) \, \epsilon_{\bar{D}^*}^\mu(P-k)\, \epsilon_{D_s^*}^{\alpha}(k) \nonumber\\
  &\quad\times \left[ -g_{\mu\nu} I_{\alpha}^1 - \frac{1}{m_{Z_c}^2}P_\nu I_{\mu \alpha}^2 + \frac{1}{m_{Z_c}^2} I_{\mu\nu\alpha}^3\right].\label{Eq:tc}
\end{align}
Note that the above expression is analogous (up to a phase) to the expression of $t_a$ in Eq.~(\ref{Eq:ta}) changing $g_L\to g_H$, $J/\psi\to \bar D^*$, $\pi\to D$, $K^*\to D^*_s$ in the couplings and in the products of four momenta, i.e., we have now $m_2=m_D$ (instead of $m_\pi$), $m_a=m_{\bar D^*}$ (instead of $m_{J/\psi}$) and $m_b=m_{D^*_s}$ (instead of $m_{K^*}$). This result is expected since we are, basically, changing a light pseudoscalar (the pion) by a heavy pseudoscalar (the $D$ meson) and light vector mesons ($J/\psi$ and $K^*$) by heavy ones ($\bar D^*$ and $D^*_s$ respectively).
\begin{figure}
\centering
\includegraphics[width=0.7\textwidth]{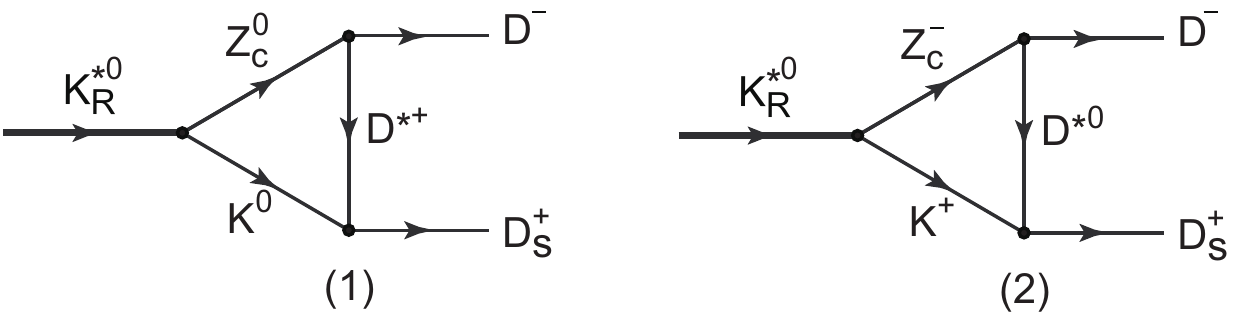}
\caption{Contributions related to the diagram (d) of Fig.~\ref{Fig:decaydiags} for the decay mechanism $K^{*0}_R\to D^{*-} D^{*+}_s$.}\label{DbarDs}
\end{figure}
 
At last, considering the vertices in Eqs.~(\ref{Eq:VKR}),~(\ref{Eq:VZcJ}),~(\ref{Eq:VKX}) and the diagrams in Fig.~\ref{DbarDs}, we get the following amplitude for the description of the process shown in Fig.~\ref{Fig:decaydiags}(d), 
\begin{align}
t_d &= t^{(1)}_d+t^{(2)}_d=i\sqrt{\frac{3}{2}}\, g_{K_R^*\to (KZ_c)_{\frac{1}{2}} }\, g_{Z_c\to(\bar{D}D^*)_{1}} \, g_H\, \epsilon_{K_R^*,\mu}(P)   \nonumber\\
&\quad\times \int\frac{d^4 q}{(2\pi)^4} \frac{\left(-g^{\mu\nu}+\frac{(P-q)^\mu (P-q)^\nu}{m_{Z_c}^2}\right)\left(-g_{\nu\alpha}+\frac{(k-q)_\nu (k-q)_\alpha}{m_{D^*}^2}\right) (k+q)^\alpha}{(q^2-m_K^2+i\epsilon) [(k-q)^2-m_{D^*}^2+i\epsilon] [(P-q)^2-m_{Z_c}^2+i\epsilon]}.\label{Eq:td}
\end{align}
Using the Lorenz gauge, we can write Eq.~(\ref{Eq:td}) as
\begin{align}	
t_d&= i\sqrt{\frac{3}{2}}\, g_{K_R^*\to (KZ_c)_{\frac{1}{2}} }\, g_{Z_c\to(\bar{D}D^*)_{1}} \, g_H\, \epsilon_{K_R^*,\mu}(P)   \nonumber\\
&\quad\times\left[ k^\mu \left(1 -\frac{k^2}{m_{D^*}^2} \right) I_0+  \left(1 + \frac{k^2}{m_{D^*}^2} + \frac{P\cdot k}{m_{Z_c}^2} - \frac{k^2 P\cdot k}{m_{D^*}^2 m_{Z_c}^2} \right)  I_1^\mu \right. + \frac{k^\mu}{m_{D^*}^2} g_{\alpha\beta}I_2^{\alpha \beta}\nonumber\\
&\quad\quad\quad +\left(\frac{(P-k)_\sigma}{m_{Z_c}^2} + \frac{k^2(P+k)_\sigma}{m_{D^*}^2 m_{Z_c}^2}  \right) I_2^{\mu\sigma}-\left( \frac{1}{m_{D^*}^2} + \frac{1}{m_{Z_c}^2} + \frac{k^2-P\cdot k}{m_{D^*}^2 m_{Z_c}^2} \right) g_{\alpha \beta} I_3^{\mu\alpha\beta}\nonumber\\
&\quad\quad\quad \left. - \frac{(P+k)_\sigma}{m_{D^*}^2 m_{Z_c}^2} g_{\alpha \beta} I_4^{\mu\alpha \beta \sigma} + \frac{1}{m_{D^*}^2 m_{Z_c}^2}g_{\alpha\beta}g_{\sigma\lambda} I_5^{\mu\alpha\beta\sigma\lambda}\right].\nonumber
\end{align}
We could proceed as in the previous cases and use the Lorentz covariance to write the tensor integrals in terms of the the possible Lorentz structures and some $a^i_j$ coefficients. However, the presence of the tensor integrals $I_4^{\mu\alpha \beta \sigma}$ and $I_5^{\mu\alpha\beta\sigma\lambda}$ makes such a method inconvenient, since many different Lorentz structures would appear. We adopt then a different strategy: although the particles in the triangular loop are off-shell, their interactions give rise to the $K^*_R$ and $Z_c$ states. In such a situation, the momenta associated with the particles generating such states are much smaller as compared to their energies. In this way, the temporal component of the polarization vector (of the order momentum/mass) is negligible as compared to the spatial components. Thus, when summing over the internal polarizations of the particles, if we call $Q^\mu$ and $m$ the four-momentum and mass, respectively, of the vector meson whose interaction with the corresponding pseudoscalar generates $K^*_R$ or $Z_c$, 
\begin{align}
-g^{\mu\nu}+\frac{Q^\mu Q^\nu}{m^2}\to\delta_{ij},\label{delta}
\end{align}
with $i$ and $j$ being spatial indices. Considering such an approach, the amplitude in Eq.~(\ref{Eq:td}) can be written as 
\begin{align}\label{Eq:tdapprox}
	t_d &= i\sqrt{\frac{3}{2}}\, g_{K_R^*\to (KZ_c)_{\frac{1}{2}} }\, g_{Z_c\to(\bar{D}D^*)_{1}} \, g_H\, \vec{\epsilon}_{K_R^*}(P)  \nonumber\\
	&\times\int\frac{d^4 q}{(2\pi)^4} \frac{\vec{k} + \vec{q}}{(q^2-m_K^2+i\epsilon) [(k-q)^2-m_{D^*}^2+i\epsilon] [(P-q)^2-m_{Z_c}^2+i\epsilon]}\nonumber\\
	&=i\sqrt{\frac{3}{2}}\, g_{K_R^*\to (KZ_c)_{\frac{1}{2}} }\, g_{Z_c\to(\bar{D}D^*)_{1}} \, g_H\, \vec{\epsilon}_{K_R^*}(P)(I^0\vec{k}+\vec{I}^{\,1}),
\end{align}
where $I^0$ is given by Eq.~(\ref{Eq:tensorInt}) (with $m_1=m_K$, $m_2=m_{D^*}$ and $m_3=m_{Z_c}$) and 
\begin{align}
\vec{I}^{\,1}\equiv\int\frac{d^4 q}{(2\pi)^4} \frac{\vec{q}}{(q^2-m_K^2+i\epsilon) [(k-q)^2-m_{D^*}^2 + i\epsilon] [(P-q)^2-m_{Z_c}^2 + i\epsilon]}.\label{vecI1}
\end{align}
Note that the approach shown in Eq.~(\ref{delta}) could have also been used when calculating the amplitudes in the diagrams depicted in Fig.~\ref{Fig:decaydiags}(a)-(c). There, however, such an approach would not lead to a significant simplification in the calculations, and we have not implemented it. In any case, for completeness, in Sect.~\ref{res}, we discuss the validity of such an approach by comparing the results obtained with and without the substitution of Eq.~(\ref{delta}) for the diagram shown in Fig.~\ref{Fig:decaydiags}(a).

The next step to get $t_d$ consists in performing the $q^0$ integration in Eq.~(\ref{vecI1}). The details of this integration are given in the appendix~\ref{ap}. After that, since in the rest frame of the decaying particle $\vec{P}=\vec{0}$, the integral in Eq.~(\ref{vecI1}) is a function of $\vec{k}$. In such a case, to perform the integration in $d^3q$, it is more convenient to introduce the dot product between $\vec{q}$ and $\vec{k}$, which can be done by replacing
\begin{equation}
	\int d^3 q\, \vec{q} \rightarrow \vec{k}\,\int d^3 q\, \frac{\vec{q}\cdot\vec{k}}{\vec{k}^2}.
\end{equation}

\section{Results and discussion}\label{res}
The decay width of the $K^*_R$ state to the two-body channels shown in Fig.~\ref{Fig:decaydiags} can be obtained from the amplitudes determined in the previous section as
\begin{equation}
	\Gamma_{i} = \int\frac{d\Omega}{4\pi^2} \frac{1}{8 M_{K_R^*}^2} \frac{p_\mathrm{c.m.}}{3}\sum |t_i|^2 = \frac{p_\mathrm{c.m.}}{24\pi M_{K_R^*}^2}\sum |t_i|^2,\label{width}
\end{equation}
where the index $i=a$, $b$, $c$, $d$ is associated with the processes shown in Fig.~\ref{Fig:decaydiags} ($a\equiv K^*_R\to J/\psi K^*$, $b\equiv K^*_R\to\bar D D^*_s$, $c\equiv K^*_R\to \bar D^* D^*_s$, $d\equiv\bar K^*_R\to \bar D D_s$), $d\Omega$ represents the solid angle, $p_\text{c.m}$ is the center of mass momentum of the particles in the final state, the factor $3$ has its origin on the average over the $K_R^*$ meson polarizations and the symbol $\sum$ indicates summation over the polarizations of the initial and final states.

Considering Eq.~(\ref{width}) and Eqs.~(\ref{Eq:tasim}),~(\ref{Eq:tb}),~(\ref{Eq:tc}),~(\ref{Eq:tdapprox}), we get, when regularizing the integrals present in the $a^i_j$ coefficients with a cut-off $\Lambda=700$ MeV,
\begin{align}
\Gamma_{a}=6.70~\text{MeV},\quad \Gamma_b=0.47~\text{MeV},\quad \Gamma_c=0.47~\text{MeV},\quad\Gamma_d=0.98~\text{MeV}.\label{Gabcd}
\end{align}
It is interesting to notice that the process depicted in Fig.~\ref{Fig:decaydiags}(b) involves an anomalous vertex~\cite{Wess:1971yu,Witten:1983tw}, the $D^*D^*_s K$ vertex, whose contribution is given by the Lagrangian in Eq.~(\ref{VVP}). It is sometimes argued that processes involving anomalous vertices should give smaller contributions that those in which no anomalous vertices are involved. However, the importance of the anomalous vertices in different contexts, like in the determination of production and absorption cross sections of several processes, calculation of radiative decays of scalar and axial resonances and kaon photo-production, has been shown~\cite{Oh:2000qr,Nagahiro:2008mn,Nagahiro:2008cv,Ozaki:2007ka,Torres:2014fxa,Abreu:2016qci,MartinezTorres:2017eio,Abreu:2017cof}. In the present work, as can be seen, the decay width found for the $\bar D D^*_s$ channel, which, as stated above, involves an anomalous vertex,  is comparable to the result obtained for the $\bar D^* D^*_s$ channel, which does not involve anomalous vertices, but has smaller phase space than $\bar D D^*_s$. 

We can study the sensitivity of the results to the cut-off used when regularizing the integrals appearing in the $a^i_j$ coefficients of Eqs.~(\ref{Eq:tasim}),~(\ref{Eq:tb}),~(\ref{Eq:tc}),~(\ref{Eq:tdapprox}).
Changing $\Lambda$ in the range 700-800 MeV, we get the following values for the decay widths
\begin{align}
\Gamma_{a}&=6.97\pm0.27~\text{MeV},\quad \Gamma_b=0.54\pm0.08~\text{MeV},\nonumber\\
\Gamma_c&=0.54\pm0.07~\text{MeV},\quad\Gamma_d=1.14\pm0.17~\text{MeV}.
\end{align}
We can also study the uncertainty produced in the results under changes in the coupling constant of $K^*_R\to KZ_c$. If we allow a variation of $\pm 1\%$ in this coupling, for a fixed cut-off $\Lambda=700$ MeV, we get
\begin{align}
\Gamma_{a}&=6.71\pm0.14~\text{MeV},\quad \Gamma_b=0.47\pm0.02~\text{MeV},\nonumber\\
\Gamma_c&=0.47\pm0.01~\text{MeV},\quad\Gamma_d=0.98\pm0.02~\text{MeV}.
\end{align}

In case of the diagram shown in Fig.~\ref{Fig:decaydiags}(a), when calculating the decay width of $K^*_R\to J/\psi K^*$, we can also consider the fact that the $K^*$ meson has a width $\Gamma_{K^*}\sim$ 47 MeV from its decay to the $K\pi$ channel. This can be done by convoluting the expression in Eq.~(\ref{width}) with the spectral function associated with the $K^*$ meson, in which case
\begin{align}
\Gamma_a=\frac{1}{N}\int\limits_{(m_{K^*}-2\Gamma_{K^*})^2}^{(m_{K^*}+2\Gamma_{K^*})^2} d\tilde{m}^2\,\text{Im}\left[\frac{1}{\tilde{m}^2-m_{K^*}^2+i \Gamma_{K^*}(\tilde{m}^2)\tilde{m}}\right]\Gamma_a(\tilde{m}^2)\Theta(m_{K^*_R}-m_{J/\psi}-\tilde{m}),\label{newG}
\end{align}
where 
\begin{align}
N=\int\limits_{(m_{K^*}-2\Gamma_{K^*})^2}^{(m_{K^*}+2\Gamma_{K^*})^2} d\tilde{m}^2\,\text{Im}\left[\frac{1}{\tilde{m}^2-m_{K^*}^2+i \Gamma_{K^*}(\tilde{m}^2)\tilde{m}}\right],
\end{align}
the expression for $\Gamma_a(\tilde{m}^2)$ in Eq.~(\ref{newG}) is given by Eq.~(\ref{width}), and
\begin{align}
\Gamma_{K^*}(\tilde{m}^2)=\Gamma_{K^*}\left[\frac{p_\text{c.m}(\tilde{m}^2,m^2_K,m^2_\pi)}{p_\text{c.m}(m^2_{K^*},m^2_K,m^2_\pi)}\right]^3.
\end{align}
Note, however, that since the mass of the $K^*_R$ resonance is far from the $J/\psi K^*$ threshold, even when the width of $K^*$ is taken into account, a significant change in the results is not expected. We indeed find almost the same value for the decay width $\Gamma_a$.

It is also interesting to establish the validity of the approach in Eq.~(\ref{delta}). If we would have considered such an approach when determining the amplitude in Eq.~(\ref{Eq:ta}), the terms related to the coefficients different to $a^1_1$ would have vanished. In such a case, we would have got for $\Gamma_a$ the value of 6.66 MeV instead of the result in Eq.~(\ref{Gabcd}). This clearly shows that the approach in Eq.~(\ref{delta}) is, in fact, reliable.

\section{Conclusion}
In this work we have calculated the decay width of the $K^*(4307)$ predicted in Ref.~\cite{Ren:2018pcd} to the two-body channels $J/\psi K^*$, $\bar D D^*_s$, $\bar D^* D^*_s$ and $\bar D D_s$. These channels, as well as the decay mechanism, are related to the internal structure of the proposed $K^*(4307)$, which, as found in Ref.~\cite{Ren:2018pcd}, corresponds to a $KD\bar D^*$ system in which the $D\bar D^*$ subsystem clusters as $X(3872)$ or $Z_c(3900)$. The possible formation of vector meson resonances with strangeness at the charmonium energy region has been, so far, unexplored. The mass and quantum numbers of the state invoke a clear non quark-antiquark structure for it. The results presented in this work constitute a prediction for the decay properties of this $K^*(4307)$ and should serve as a motivation for conducting experimental investigations of this state.

\section*{Acknowledgements}
This work was partly supported by DFG and NSFC through funds provided to the Sino-German CRC 110 ``Symmetries and the Emergence of Structure in QCD'' (Grant No. TRR110), CNPq (Grant No. 310759/2016-1 and 311524/2016-8), and NSFC (Grant No. 11775099). X.~-L.~Ren  thanks to the valuable discussion with Profs.~Cheng-Ping Shen and Li-Ming Zhang during the QNP 2018 conference, which motived the present calculation.
\appendix
\section{Determination of the $a^i_j$ coefficients involved in the triangular loops depicted in Fig.~\ref{Fig:decaydiags}}\label{ap}
In this appendix we give the details of the calculation of the $a^i_j$ coefficients appearing in Eqs.~(\ref{Eq:ta}),~(\ref{Eq:tb}),~(\ref{Eq:tc}), which are related to the Lorentz decomposition of Eq.~(\ref{Is}),
and to determine Eq.~(\ref{vecI1}). By contracting $I_\alpha^1$, $I_{\mu\alpha}^2$
and $I_{\mu\nu\alpha}^3$ with the corresponding Lorentz structures, we can get a set of coupled equations whose solution, in each case, allow us to write the $a^i_j$ coefficients in terms of scalar integrals. We obtain (the expression for the $a^1_1$ coefficient can be found in Eq.~(\ref{a11}) but, for convenience, we write it here again)
\begin{align}
a_{1}^1&= \frac{k^2(\mathbb{PI}^1)-(k\cdot P)(\mathbb{KI}^1)}{k^2 P^2 - (k\cdot P)^2},\nonumber\\
a_1^2&= \frac{\left[(P\cdot k)^2-k^2 P^2\right] \mathbb{G I}^2 + k^2 \mathbb{PPI}^2 + P^2 \mathbb{KKI}^2 - 2(P\cdot k) \mathbb{KPI}^2}{2[(P\cdot k)^2 - k^2 P^2]},\nonumber\\
a_3^2 &= \frac{1}{2[(P\cdot k)^2 - k^2 P^2]^2} \left[ P\cdot k \left(k^2P^2-(P\cdot k)^2\right) \mathbb{G I}^2 - 3 k^2 (P\cdot k) \mathbb{PPI}^2 \right. \nonumber\\
&\quad \left. + 2\left(k^2 P^2 + 2(P\cdot k)^2\right) \mathbb{KPI}^2 - 3 P^2 (P\cdot k) \mathbb{KKI}^2 \right], \nonumber\\
a_1^3 &= \frac{1}{2\left[(P\cdot k)^2 - k^2 P^2\right]^2} \left[k^2\left(k^2 P^2 - (P\cdot k)^2\right)  \mathbb{GPI}^3- P\cdot k\left(k^2 P^2 - (P\cdot k)^2 \right) \mathbb{GKI}^3 \right.\nonumber\\
&\quad- k^4 \mathbb{PPPI}^3 + 3 k^2 (P\cdot k) \mathbb{PPKI}^3 + P^2 (P\cdot k) \mathbb{KKKI}^3\left. - \left(k^2 P^2 + 2(P\cdot k)^2\right) \mathbb{KKPI}^3 \right],\nonumber\\
a_2^3 &= \frac{1}{2\left[(P\cdot k)^2 - k^2 P^2\right]^2} \left[ - P\cdot k \left(k^2 P^2 - (P\cdot k)^2\right) \mathbb{GPI}^3 \right.+ P^2 \left(k^2 P^2 - (P\cdot k)^2 \right) \mathbb{GKI}^3 \nonumber\\
&\quad+ k^2 (P\cdot k) \mathbb{PPPI}^3 - \left(k^2 P^2 + 2(P\cdot k)^2 \right) \mathbb{PPKI}^3 - P^4 \mathbb{KKKI}^3\left. + 3 P^2 (P\cdot k) \mathbb{KKPI}^3 \right],\label{aCs}\\
 a_4^3 &= -\frac{1}{2\left[(P\cdot k)^2 - k^2 P^2\right]^3} \left[ 3 k^2 (P\cdot k) \left(k^2 P^2 - (P\cdot k)^2\right) \mathbb{GPI}^3 \right.- \left(k^2 P^2 - (P\cdot k)^2 \right)\nonumber\\
 &\quad\times\left(k^2 P^2 + 2 (P\cdot k)^2 \right) \mathbb{GKI}^3 -5 k^4 (P\cdot k) \mathbb{PPPI}^3 + 
3k^2\left(k^2 P^2 + 4 (P\cdot k)^2\right) \mathbb{PPKI}^3 \nonumber\\
& \quad + P^2\left(k^2 P^2 + 4(P\cdot k)^2\right) \mathbb{KKKI}^3\left. - 3(P\cdot k) \left(3k^2P^2+2(P\cdot k)^2\right) \mathbb{KKPI}^3\right],\nonumber\\
a_{6}^3 &= -\frac{1}{2\left[(P\cdot k)^2 - k^2 P^2\right]^3}
\left[ \left((P\cdot k)^2 - k^2 P^2\right) \left(2(P\cdot k)^2 + k^2 P^2\right) \mathbb{GPI}^3 \right. \nonumber\\ 
&\quad -3 P^2 (P\cdot k)\left((P\cdot k)^2-k^2 P^2\right) \mathbb{GKI}^3 +  k^2 \left(4(P\cdot k)^2 + k^2 P^2\right) \mathbb{PPPI}^3 \nonumber\\
&\quad -3 (P\cdot k)\left(2(P\cdot k)^2 + 3 k^2 P^2 \right) \mathbb{PPK I}^3 -5 P^4 (P\cdot k) \mathbb{KKKI}^3 \nonumber\\
& \quad \left. + 3P^2\left(4(P\cdot k)^2 + k^2 P^2\right) \mathbb{KKPI}^3 \right].\nonumber
\end{align}
where,
\begin{align}
&\mathbb{PI}^1=P^\mu I^1_\mu,\quad\mathbb{KI}^1=k^\mu I^1_\mu,\nonumber\\
&\mathbb{GI}^2\equiv g^{\mu\alpha} I_{\mu\alpha}^2,\quad\mathbb{PPI}^2\equiv P^\mu P^\alpha I_{\mu\alpha}^2,\quad \mathbb{KPI}^2\equiv k^\mu P^{\alpha} I_{\mu\alpha}^2,\quad
\mathbb{KKI}^2\equiv k^\mu k^\alpha I_{\mu\alpha}^2,\nonumber\\
&\mathbb{GPI}^3\equiv g^{\mu\nu} P^{\alpha} I_{\mu\nu\alpha}^3,\quad\mathbb{GKI}^3\equiv g^{\mu\nu} k^{\alpha} I_{\mu\nu\alpha}^3,\quad\mathbb{PPPI}^3\equiv P^\mu P^\nu P^\alpha I_{\mu\nu\alpha}^3,\label{Isca}\\
&\mathbb{PPKI}^3\equiv P^\mu P^\nu k^\alpha I_{\mu\nu\alpha}^3,\quad\mathbb{KKKI}^3\equiv k^\mu k^\nu k^\alpha I_{\mu\nu\alpha}^3,\quad\mathbb{KKPI}^3\equiv k^\mu k^\nu P^\alpha I_{\mu\nu\alpha}^3.\nonumber
\end{align}
Note that the $a^i_j$ coefficients in Eq.~(\ref{aCs}) and the scalars in Eq.~(\ref{Isca}) depend on the masses $m_1$, $m_2$ and $m_3$ of the particles involved in the triangular loop as well as of the mass of the $K^*_R$  state and the masses of the particles to which it can decay, which we represent by $m_a$ and $m_b$ (see Fig.~\ref{tri}). 

Using Eq.~(\ref{Eq:tensorInt}), and working in the rest frame of the decaying particle, we can write
\begin{align}
&\mathbb{PI}^1=\int\frac{dq^0}{(2\pi)}\int\frac{d^3q}{(2\pi)^3}\frac{P^0 q^0}{F(q^0,\vec{q})},\quad \mathbb{KI}^1=\int\frac{dq^0}{(2\pi)}\int\frac{d^3q}{(2\pi)^3}\frac{k^0q^0-\vec{k}\cdot\vec{q}}{F(q^0,\vec{q})},\nonumber\\
&\mathbb{GI}^2=\int\frac{dq^0}{(2\pi)}\int\frac{d^3q}{(2\pi)^3}\frac{{q^0}^2-\vec{q}^{\,2}}{F(q^0,\vec{q})},\quad
\mathbb{PPI}^2=\int\frac{dq^0}{(2\pi)}\int\frac{d^3q}{(2\pi)^3}\frac{(P^0q^0)^2}{F(q^0,\vec{q})},\label{sc1}\\
&\mathbb{KPI}^2=\int\frac{dq^0}{(2\pi)}\int\frac{d^3q}{(2\pi)^3}\frac{(k^0q^0-\vec{k}\cdot\vec{q})P^0q^0}{F(q^0,\vec{q})},\quad
\mathbb{KKI}^2=\int\frac{dq^0}{(2\pi)}\int\frac{d^3q}{(2\pi)^3}\frac{(k^0 q^0-\vec{k}\cdot\vec{q})^2}{F(q^0,\vec{q})}.\nonumber
\end{align}
where 
\begin{align}
F(q^0,\vec{q})=[{q^0}^2-\omega^2_1+i\epsilon][(k^0-q^0)^2-\omega^2_2+i\epsilon][(P^0-q^0)^2-\omega^2_3+i\epsilon],
\end{align}
with $\omega_1=\sqrt{\vec{q}^{\,\,2}+m^2_1}$, $\omega_2=\sqrt{(\vec{k}-\vec{q})^2+m^2_2}$ and $\omega_3=\sqrt{\vec{q}^{\,\,2}+m^2_3}$. The integrals in Eq.~(\ref{sc1}) are particular cases of the most general integral $I(a,b,b^\prime,c,d,e)$ defined as
\begin{align}
I(a,b,b^\prime,c,d,e)=\int\frac{dq^0}{(2\pi)}\int\frac{d^3q}{(2\pi)^3}\frac{a\, {q^0}^2+(b+b^\prime{\text{cos}^2\theta})\vec{q}^{\,2}+c\, q^0|\vec{q}|\text{cos}\theta+d\,q^0+e\,|\vec{q}|\text{cos}\theta}{[{q^0}^2-\omega^2_1+i\epsilon][(k^0-q^0)^2-\omega^2_2+i\epsilon][(P^0-q^0)^2-\omega^2_3+i\epsilon]}.\label{Igen}
\end{align}
Indeed, we can write the integrals in Eq.~(\ref{sc1}) as
\begin{align}
&\mathbb{PI}^1=I(0,0,0,0,P^0,0),\quad\mathbb{KI}^1=I(0,0,0,0,k^0,-|\vec{k}|),\quad\mathbb{GI}^2=I(1,-1,0,0,0,0)\nonumber\\
&\mathbb{PPI}^2=I({P^0}^2,0,0,0,0,0),\quad\mathbb{KPI}^2=I(k^0 P^0,0,0,-|\vec{k}|P^0,0,0),\nonumber\\
&\mathbb{KKI}^2=I({k^0}^2,0,|\vec{k}|^2,-2k^0|\vec{k}|,0,0),
\end{align}
where $P^0=m_{K^*_R}$ and 
\begin{align}
k^0=\frac{{P^0}^2-m^2_a+m^2_b}{2P^0},\quad|\vec{k}|=\frac{\lambda^{1/2}({P^0}^2,m^2_a,m^2_b)}{2P^0}.
\end{align}
The $q^0$ integration in Eq.~(\ref{Igen}) can be performed analytically by using Cauchy's theorem. After that, the resulting integration in $d^3q$ is regularized. This is done by means of a cut-off $\Lambda\sim 700$ MeV, in agreement with the cut-off used in the study of the $KD\bar D^*$ system, in which the $K^*_R$ state was predicted~\cite{Ren:2018pcd}. In this way, we get
\begin{align}
I(a,b,b^\prime,c,d,e)=-\frac{i}{(2\pi)^2}\int\limits_0^\Lambda dq q^2\int\limits_{-1}^1d\text{cos}\theta\,\frac{N(q,\theta;a,b,b^\prime,c,d,e)}{D(q,\theta)},
\end{align}
with $q=|\vec{q}|$ and
\begin{align}
N&(q,\theta;a,b,b^\prime,c,d,e)=a\,\omega_1\Big[{k^0}^2\{\omega_3(\omega_1+\omega_3)(\omega_1+\omega_2+\omega_3)-{P^0}^2(\omega_2+\omega_3)\}\nonumber\\
&+2 k^0 P^0\omega_1\omega_2\omega_3-\omega_2(\omega_1+\omega_2)\{\omega_3(\omega_1+\omega_3)(\omega_2+\omega_3)-{P^0}^2(\omega_1+\omega_2+\omega_3)\}\Big]\nonumber\\
&+q(\tilde{b}q+\tilde{e})\Big[-{k^0}^2\omega_2(\omega_1+\omega_3)+2 k^0 P^0\omega_2\omega_3+(\omega_1+\omega_2)\{(\omega_1+\omega_3)(\omega_2+\omega_3)\nonumber\\
&\quad\quad\times(\omega_1+\omega_2+\omega_3)-{P^0}^2\omega_3\}\Big]+(\tilde{c}q+d)\omega_1\Big[-{k^0}^2 P^0\omega_2+k^0\omega_3\{(\omega_1+\omega_3)\nonumber\\
&\quad\quad\times(\omega_1+2\omega_2+\omega_3)-{P^0}^2\}+P^0\omega_2(\omega_1+\omega_2)(\omega_1+\omega_2+2\omega_3)\Big],\nonumber\\
D&(q,\theta)=2\omega_1\omega_2\omega_3(k^0-\omega_1-\omega_2+i\epsilon)(k^0+\omega_1+\omega_2)(P^0-\omega_1-\omega_3+i\epsilon)\nonumber\\
&\quad\quad\quad\times(P^0+\omega_1+\omega_3)(P^0-k^0-\omega_2-\omega_3+i\epsilon)(k^0-P^0-\omega_2-\omega_3+i\epsilon),
\end{align}
where we have introduced $\tilde{b}\equiv b+b^\prime\text{cos}^2\theta$, $\tilde{c}=c\,\text{cos}\theta$ and $\tilde{e}=e\,\text{cos}\theta$.

Similarly, 
\begin{align}
\mathbb{GPI}^3&=\int\frac{dq^0}{(2\pi)}\int\frac{d^3q}{(2\pi)^3}\frac{({q^0}^2-|\vec{q}|^2)(P^0 q^0)}{F(q^0,\vec{q})},\nonumber\\
\mathbb{GKI}^3&=\int\frac{dq^0}{(2\pi)}\int\frac{d^3q}{(2\pi)^3}\frac{({q^0}^2-|\vec{q}|^2)(k^0 q^0-|\vec{k}||\vec{q}|\text{cos}\theta)}{F(q^0,\vec{q})},\nonumber\\
\mathbb{PPPI}^3&=\int\frac{dq^0}{(2\pi)}\int\frac{d^3q}{(2\pi)^3}\frac{(P^0q^0)^3}{F(q^0,\vec{q})},\nonumber\\
\mathbb{PPKI}^3&=\int\frac{dq^0}{(2\pi)}\int\frac{d^3q}{(2\pi)^3}\frac{(P^0q^0)^2(k^0q^0-|\vec{k}||\vec{q}|\text{cos}\theta)}{F(q^0,\vec{q})},\label{Isca3}\\
\mathbb{KKKI}^3&=\int\frac{dq^0}{(2\pi)}\int\frac{d^3q}{(2\pi)^3}\frac{(k^0q^0-|\vec{k}||\vec{q}|\text{cos}\theta)^3}{F(q^0,\vec{q})},\nonumber\\
\mathbb{KKPI}^3&=\int\frac{dq^0}{(2\pi)}\int\frac{d^3q}{(2\pi)^3}\frac{(k^0q^0-|\vec{k}||\vec{q}|\text{cos}\theta)^2P^0q^0}{F(q^0,\vec{q})}.\nonumber
\end{align}
The integrals in Eq.~(\ref{Isca3}) are particular cases of the most general integral $\mathcal{I}(a,b,c,d,e,e^\prime,f,f^\prime,f^{\prime\prime})$ defined as
\begin{align}
\mathcal{I}(a&,b,c,d,e,e^\prime,f,f^\prime,f^{\prime\prime})\nonumber\\
&=\int\frac{dq^0}{(2\pi)}\int\frac{d^3q}{(2\pi)^3}\frac{a\,{q^0}^4+\tilde{b}\,{q^0}^3|\vec{q}|+c\,{q^0}^3+\tilde{d}{q^0}^2|\vec{q}|+\tilde{e}q^0|\vec{q}|^2+\tilde{f}|\vec{q}|^3}{[{q^0}^2-\omega^2_1+i\epsilon][(k^0-q^0)^2-\omega^2_2+i\epsilon][(P^0-q^0)^2-\omega^2_3+i\epsilon]},\label{Ical}
\end{align}
with $\tilde{b}\equiv b\,\text{cos}\theta$, $\tilde{d}\equiv d\,\text{cos}\theta$, $\tilde{e}\equiv(e+e^\prime\text{cos}^2\theta)$, $\tilde{f}\equiv(f+f^\prime\text{cos}\theta+f^{\prime\prime}\text{cos}^3\theta)$. Particularly, we can write
\begin{align}
&\mathbb{GPI}^3=\mathcal{I}(0,0,P^0,0,-P^0,0,0,0,0),\quad\mathbb{GKI}^3=\mathcal{I}(0,0,k^0,-|\vec{k}|,-k^0,0,0,|\vec{k}|,0),\nonumber\\
&\mathbb{PPI}^3=\mathcal{I}(0,0,{P^0}^3,0,0,0,0,0,0),\quad \mathbb{PPKI}^3=\mathcal{I}(0,0,{P^0}^2k^0,-{P^0}^2|\vec{k}|,0,0,0,0,0),\nonumber\\
&\mathbb{KKKI}^3=\mathcal{I}(0,0,{k^0}^3,-3{k^0}^2|\vec{k}|,0,3k^0|\vec{k}|^2,0,0,-|\vec{k}|^3),\\
&\mathbb{KKPI}^3=\mathcal{I}(0,0,{k^0}^2P^0,-2k^0|\vec{k}|P^0,0,|\vec{k}|^2P^0,0,0,0).\nonumber
\end{align}
The integral on the $q^0$ variable of Eq.~(\ref{Ical}) can be obtained using Cauchy's theorem and the remaining integration in $d^3q$ is regularized using a cut-off $\Lambda\sim 700$ MeV. In this way,
\begin{align}
\mathcal{I}(a,b,c,d,&e,e^\prime,f,f^\prime,f^{\prime\prime})=-\frac{i}{(2\pi)^2}\int\limits_0^\Lambda dq q^2\int\limits_{-1}^{1}d\text{cos}\theta
\,\frac{\mathcal{N}(q,\theta;a,b,c,d,e,e^\prime,f,f^\prime,f^{\prime\prime})}{D(q,\theta)},
\end{align}
where 
\begin{align}
\mathcal{N}(q,\theta;&a,b,c,d,e,e^\prime,f,f^\prime,f^{\prime\prime})=a\,\omega_1\Big[{k^0}^4\omega_3\{(\omega_1+\omega_3)^2-{P^0}^2\}\nonumber\\
&-{k^0}^2\omega_2\Big\{{P^0}^4-2{P^0}^2\omega_3(2\omega_1+\omega_2+\omega_3)+\omega_3(\omega_1+\omega_3)(\omega_1^2+2\omega_1\omega_2+3\omega_1\omega_3\nonumber\\
&+2\omega_2\omega_3+\omega^2_3)\Big\}+2k^0 P^0\omega^3_1\omega_2\omega_3+\omega_2(\omega_1+\omega_2)\Big\{{P^0}^4(\omega_1+\omega_2)\nonumber\\
&-{P^0}^2\omega_3\Big(\omega^2_1+2\omega_3(\omega_1+\omega_2)+3\omega_1\omega_2+\omega^2_2\Big)+\omega_3(\omega_1+\omega_3)(\omega_2+\omega_3)\nonumber\\
&\times\Big(\omega_1(\omega_2+\omega_3)+\omega_2\omega_3\Big)\Big\}\Big]+q\Big[\tilde{b}\,\omega_1\Big\{{k^0}^3\omega_3\Big((\omega_1+\omega_3)^2-{P^0}^2\Big)\nonumber\\
&+{k^0}^2 P^0\omega_2\Big(\omega_3(2\omega_1+\omega_3)-{P^0}^2\Big)-k^0\omega_2\omega_3\Big((\omega_1+\omega_3)\{\omega_1(\omega_2+2\omega_3)\nonumber\\
&+\omega_2\omega_3\}-{P^0}^2(2\omega_1+\omega_2)\Big)+P^0\omega_2(\omega_1+\omega_2)\Big({P^0}^2(\omega_1+\omega_2)\nonumber\\
&-\omega_3\{\omega_3(\omega_1+\omega_2)+2\omega_1\omega_2\}\Big)\Big\}+\tilde{d}\,\omega_1\Big\{{k^0}^2\Big(\omega_3(\omega_1+\omega_3)(\omega_1+\omega_2+\omega_3)\nonumber\\
&-{P^0}^2(\omega_2+\omega_3)\Big)+2k^0P^0\omega_1\omega_2\omega_3-\omega_2(\omega_1+\omega_2)\Big(\omega_3(\omega_1+\omega_3)(\omega_2+\omega_3)\nonumber\\
&-{P^0}^2(\omega_1+\omega_2+\omega_3)\Big)\Big\}+q\Big\{\tilde{e}\,\omega_1\Big(-{k^0}^2P^0\omega_2+k^0\omega_3\{(\omega_1+\omega_3)(\omega_1+2\omega_2+\omega_3)\nonumber\\
&-{P^0}^2\}+P^0\omega_2(\omega_1+\omega_2)(\omega_1+\omega_2+2\omega_3)\Big)+\tilde{f}q\Big(-{k^0}^2\omega_2(\omega_1+\omega_3)+2k^0P^0\omega_2\omega_3\nonumber\\
&+(\omega_1+\omega_2)\{(\omega_1+\omega_3)(\omega_2+\omega_3)(\omega_1+\omega_2+\omega_3)-{P^0}^2\omega_3\}\Big)\Big\}\Big]\nonumber\\
&-c\,\omega_1\Big[{k^0}^3\omega_3(P^0-\omega_1-\omega_3)(P^0+\omega_1+\omega_3)+{k^0}^2P^0\omega_2\{{P^0}^2-\omega_3(2\omega_1+\omega_3)\}\nonumber\\
&+k^0\omega_2\omega_3\{(\omega_1+\omega_3)(\omega_1\omega_2+2\omega_1\omega_3+\omega_2\omega_3)-{P^0}^2(2\omega_1+\omega_2)\}\nonumber\\
&+P^0\omega_2(\omega_1+\omega_2)\{\omega_3(\omega_3(\omega_1+\omega_2)+2\omega_1\omega_2\big)-{P^0}^2(\omega_1+\omega_2)\}\Big].
\end{align}
Next, we determine the $I^0$ and $\vec{I}^{\,1}$ integrals of Eq.~(\ref{Eq:tdapprox}). By means of the Cauchy's theorem we can integrate on the $q^0$ variable and get the following integration in $d^3q$, which is regularized by using a cut-off $\Lambda\sim 700$ MeV,
\begin{align}
I^0&=-\frac{i}{(2\pi)^2}\int\limits_0^\Lambda dq q^2\int\limits_{-1}^{1}d\text{cos}\theta\frac{\mathbb{N}(q,\theta)}{D(q,\theta)},\nonumber\\
\vec{I}^{\,1}&=-\frac{i}{(2\pi)^2|\vec{k}|^2}\vec{k}\int\limits_0^\Lambda dq q^2\int\limits_{-1}^{1}d\text{cos}\theta\frac{\mathbb{N}(q,\theta)}{D(q,\theta)}\vec{k}\cdot\vec{q},
\end{align}
where
\begin{align}
\mathbb{N}(q,\theta)&=-{k^0}^2\omega_2(\omega_1+\omega_3)+2k^0 P^0\omega_2\omega_3+(\omega_2+\omega_3)\nonumber\\
&\quad\times\Big[(\omega_2+\omega_3)(\omega_1+\omega_3)(\omega_1+\omega_2+\omega_3)-{P^0}^2\omega_3\Big].
\end{align}

\bibliographystyle{unsrt}
\bibliography{refs}

\begin{thebibliography}{10}

\bibitem{Jaffe:1976ig}
Robert~L. Jaffe.
\newblock {Multi-Quark Hadrons. 1. The Phenomenology of (2 Quark 2 anti-Quark)
  Mesons}.
\newblock {\em Phys. Rev.}, D15:267, 1977.

\bibitem{Weinstein:1982gc}
John~D. Weinstein and Nathan Isgur.
\newblock {Do Multi-Quark Hadrons Exist?}
\newblock {\em Phys. Rev. Lett.}, 48:659, 1982.

\bibitem{vanBeveren:1986ea}
E.~van Beveren, T.~A. Rijken, K.~Metzger, C.~Dullemond, G.~Rupp, and J.~E.
  Ribeiro.
\newblock {A Low Lying Scalar Meson Nonet in a Unitarized Meson Model}.
\newblock {\em Z. Phys.}, C30:615--620, 1986.

\bibitem{Tornqvist:1995kr}
Nils~A. Tornqvist.
\newblock {Understanding the scalar meson q anti-q nonet}.
\newblock {\em Z. Phys.}, C68:647--660, 1995.

\bibitem{Oller:1997ng}
J.~A. Oller, E.~Oset, and J.~R. Pelaez.
\newblock {Nonperturbative approach to effective chiral Lagrangians and meson
  interactions}.
\newblock {\em Phys. Rev. Lett.}, 80:3452--3455, 1998.

\bibitem{Oller:1998hw}
J.~A. Oller, E.~Oset, and J.~R. Pelaez.
\newblock {Meson meson interaction in a nonperturbative chiral approach}.
\newblock {\em Phys. Rev.}, D59:074001, 1999.
\newblock [Erratum: Phys. Rev.D75,099903(2007)].

\bibitem{Dalitz:1959dn}
R.~H. Dalitz and S.~F. Tuan.
\newblock {A possible resonant state in pion-hyperon scattering}.
\newblock {\em Phys. Rev. Lett.}, 2:425--428, 1959.

\bibitem{Dalitz:1960du}
R.~H. Dalitz and S.~F. Tuan.
\newblock {The phenomenological description of -K -nucleon reaction processes}.
\newblock {\em Annals Phys.}, 10:307--351, 1960.

\bibitem{Kaiser:1995eg}
Norbert Kaiser, P.~B. Siegel, and W.~Weise.
\newblock {Chiral dynamics and the low-energy kaon - nucleon interaction}.
\newblock {\em Nucl. Phys.}, A594:325--345, 1995.

\bibitem{Oset:1997it}
E.~Oset and A.~Ramos.
\newblock {Nonperturbative chiral approach to s wave anti-K N interactions}.
\newblock {\em Nucl. Phys.}, A635:99--120, 1998.

\bibitem{Meissner:1999vr}
Ulf-G. Mei{\ss}ner and J.~A. Oller.
\newblock {Chiral unitary meson baryon dynamics in the presence of resonances:
  Elastic pion nucleon scattering}.
\newblock {\em Nucl. Phys.}, A673:311--334, 2000.

\bibitem{Klempt:2007cp}
Eberhard Klempt and Alexander Zaitsev.
\newblock {Glueballs, Hybrids, Multiquarks. Experimental facts versus QCD
  inspired concepts}.
\newblock {\em Phys. Rept.}, 454:1--202, 2007.

\bibitem{Brambilla:2010cs}
N.~Brambilla et~al.
\newblock {Heavy quarkonium: progress, puzzles, and opportunities}.
\newblock {\em Eur. Phys. J.}, C71:1534, 2011.

\bibitem{Hosaka:2016pey}
Atsushi Hosaka, Toru Iijima, Kenkichi Miyabayashi, Yoshihide Sakai, and
  Shigehiro Yasui.
\newblock {Exotic hadrons with heavy flavors: X, Y, Z, and related states}.
\newblock {\em PTEP}, 2016(6):062C01, 2016.

\bibitem{Oset:2016lyh}
Eulogio Oset et~al.
\newblock {Weak decays of heavy hadrons into dynamically generated resonances}.
\newblock {\em Int. J. Mod. Phys.}, E25:1630001, 2016.

\bibitem{Lebed:2016hpi}
Richard~F. Lebed, Ryan~E. Mitchell, and Eric~S. Swanson.
\newblock {Heavy-Quark QCD Exotica}.
\newblock {\em Prog. Part. Nucl. Phys.}, 93:143--194, 2017.

\bibitem{Chen:2016qju}
Hua-Xing Chen, Wei Chen, Xiang Liu, and Shi-Lin Zhu.
\newblock {The hidden-charm pentaquark and tetraquark states}.
\newblock {\em Phys. Rept.}, 639:1--121, 2016.

\bibitem{Olsen:2017bmm}
Stephen~Lars Olsen, Tomasz Skwarnicki, and Daria Zieminska.
\newblock {Nonstandard heavy mesons and baryons: Experimental evidence}.
\newblock {\em Rev. Mod. Phys.}, 90(1):015003, 2018.

\bibitem{Guo:2017jvc}
Feng-Kun Guo, Christoph Hanhart, Ulf-G. Mei{\ss}ner, Qian Wang, Qiang Zhao, and
  Bing-Song Zou.
\newblock {Hadronic molecules}.
\newblock {\em Rev. Mod. Phys.}, 90(1):015004, 2018.

\bibitem{Liu:2019zoy}
Yan-Rui Liu, Hua-Xing Chen, Wei Chen, Xiang Liu, and Shi-Lin Zhu.
\newblock {Pentaquark and Tetraquark states}.
\newblock 2019.

\bibitem{Tanabashi:2018oca}
M.~Tanabashi et~al.
\newblock {Review of Particle Physics}.
\newblock {\em Phys. Rev.}, D98(3):030001, 2018.

\bibitem{Armstrong:1982tw}
T.~Armstrong et~al.
\newblock {A Partial Wave Analysis of the $K^- \phi$ System Produced in the
  Reaction $K^- p \to K^+ K^- K^- p$ at 18.5-{GeV}/$c$}.
\newblock {\em Nucl. Phys.}, B221:1--15, 1983.

\bibitem{Aaij:2016iza}
Roel Aaij et~al.
\newblock {Observation of $J/\psi\phi$ structures consistent with exotic states
  from amplitude analysis of $B^+\to J/\psi \phi K^+$ decays}.
\newblock {\em Phys. Rev. Lett.}, 118(2):022003, 2017.

\bibitem{Estabrooks:1977xe}
P.~Estabrooks, R.~K. Carnegie, Alan~D. Martin, W.~M. Dunwoodie, T.~A. Lasinski,
  and David W. G.~S. Leith.
\newblock {Study of K pi Scattering Using the Reactions K+- p ---> K+- pi+ n
  and K+- p ---> K+- pi- Delta++ at 13-GeV/c}.
\newblock {\em Nucl. Phys.}, B133:490--524, 1978.

\bibitem{Etkin:1980me}
A.~Etkin et~al.
\newblock {MEASUREMENT AND PARTIAL WAVE ANALYSIS OF THE REACTION K- P --->
  K0(S) PI+ PI- N AT 6-GEV/C}.
\newblock {\em Phys. Rev.}, D22:42--60, 1980.

\bibitem{Aston:1986jb}
D.~Aston et~al.
\newblock {The Strange Meson Resonances Observed in the Reaction $K^- p \to
  \bar{K}0 \pi^+ \pi^- n$ at 11-{GeV}/$c$}.
\newblock {\em Nucl. Phys.}, B292:693, 1987.

\bibitem{Aston:1987ir}
D.~Aston et~al.
\newblock {A Study of K- pi+ Scattering in the Reaction K- p ---> K- pi+ n at
  11-GeV/c}.
\newblock {\em Nucl. Phys.}, B296:493--526, 1988.

\bibitem{Ma:2018vhp}
Li~Ma, Qian Wang, and Ulf-G. Mei{\ss}ner.
\newblock {Tri-meson bound state $BBB^{\ast}$ via delocalized $\pi$~Bond}.
\newblock 2018.

\bibitem{Ren:2018pcd}
Xiu-Lei Ren, Brenda~B. Malabarba, Li-Sheng Geng, K.~P. Khemchandani, and
  A.~Mart\'inez~Torres.
\newblock {$K^*$ mesons with hidden charm arising from $KX(3872)$ and
  $KZ_c(3900)$ dynamics}.
\newblock {\em Phys. Lett.}, B785:112--117, 2018.

\bibitem{Kamalov:2000iy}
S.~S. Kamalov, E.~Oset, and A.~Ramos.
\newblock {Chiral unitary approach to the K- deuteron scattering length}.
\newblock {\em Nucl. Phys.}, A690:494--508, 2001.

\bibitem{Xie:2010ig}
Ju-Jun Xie, A.~Martinez~Torres, and E.~Oset.
\newblock {Faddeev fixed center approximation to the $N\bar{K}K$ system and the
  signature of a $N^*(1920)(1/2^+)$ state}.
\newblock {\em Phys. Rev.}, C83:065207, 2011.

\bibitem{Roca:2010tf}
L.~Roca and E.~Oset.
\newblock {A description of the f2(1270), rho3(1690), f4(2050), rho5(2350) and
  f6(2510) resonances as multi-rho(770) states}.
\newblock {\em Phys. Rev.}, D82:054013, 2010.

\bibitem{MartinezTorres:2010ax}
A.~Martinez~Torres, E.~J. Garzon, E.~Oset, and L.~R. Dai.
\newblock {Limits to the Fixed Center Approximation to Faddeev equations: the
  case of the $\phi(2170)$}.
\newblock {\em Phys. Rev.}, D83:116002, 2011.

\bibitem{Bayar:2011qj}
M.~Bayar, J.~Yamagata-Sekihara, and E.~Oset.
\newblock {The $\bar{K}NN$ system with chiral dynamics}.
\newblock {\em Phys. Rev.}, C84:015209, 2011.

\bibitem{Bayar:2015oea}
M.~Bayar, Xiu-Lei Ren, and E.~Oset.
\newblock {States of $\rho D^* \bar D^*$ with J = 3 within the fixed center
  approximation to the Faddeev equations}.
\newblock {\em Eur. Phys. J.}, A51(5):61, 2015.

\bibitem{Debastiani:2017vhv}
V.~R. Debastiani, J.~M. Dias, and E.~Oset.
\newblock {Study of the $DKK$ and $DK\bar{K}$ systems}.
\newblock {\em Phys. Rev.}, D96(1):016014, 2017.

\bibitem{Ren:2018qhr}
Xiu-Lei Ren and Zhi-Feng Sun.
\newblock {Possible bound states with hidden bottom from
  $\bar{K}^{(*)}B^{(*)}\bar{B}^{(*)}$ systems}.
\newblock 2018.

\bibitem{Gasser:1983yg}
J.~Gasser and H.~Leutwyler.
\newblock {Chiral Perturbation Theory to One Loop}.
\newblock {\em Annals Phys.}, 158:142, 1984.

\bibitem{Gasser:1984gg}
J.~Gasser and H.~Leutwyler.
\newblock {Chiral Perturbation Theory: Expansions in the Mass of the Strange
  Quark}.
\newblock {\em Nucl. Phys.}, B250:465--516, 1985.

\bibitem{Voloshin:1978hc}
M.~B. Voloshin.
\newblock {On Dynamics of Heavy Quarks in Nonperturbative QCD Vacuum}.
\newblock {\em Nucl. Phys.}, B154:365--380, 1979.

\bibitem{Isgur:1989vq}
Nathan Isgur and Mark~B. Wise.
\newblock {Weak Decays of Heavy Mesons in the Static Quark Approximation}.
\newblock {\em Phys. Lett.}, B232:113--117, 1989.

\bibitem{Burdman:1992gh}
Gustavo Burdman and John~F. Donoghue.
\newblock {Union of chiral and heavy quark symmetries}.
\newblock {\em Phys. Lett.}, B280:287--291, 1992.

\bibitem{Gamermann:2006nm}
D.~Gamermann, E.~Oset, D.~Strottman, and M.~J. Vicente~Vacas.
\newblock {Dynamically generated open and hidden charm meson systems}.
\newblock {\em Phys. Rev.}, D76:074016, 2007.

\bibitem{Guo:2006fu}
Feng-Kun Guo, Peng-Nian Shen, Huan-Ching Chiang, Rong-Gang Ping, and Bing-Song
  Zou.
\newblock {Dynamically generated 0+ heavy mesons in a heavy chiral unitary
  approach}.
\newblock {\em Phys. Lett.}, B641:278--285, 2006.

\bibitem{Nieves:2012tt}
J.~Nieves and M.~Pavon Valderrama.
\newblock {The Heavy Quark Spin Symmetry Partners of the X(3872)}.
\newblock {\em Phys. Rev.}, D86:056004, 2012.

\bibitem{Aceti:2014uea}
F.~Aceti, M.~Bayar, E.~Oset, A.~Martinez~Torres, K.~P. Khemchandani,
  Jorgivan~Morais Dias, F.~S. Navarra, and M.~Nielsen.
\newblock {Prediction of an $I=1$ $D \bar D^*$ state and relationship to the
  claimed $Z_c(3900)$, $Z_c(3885)$}.
\newblock {\em Phys. Rev.}, D90(1):016003, 2014.

\bibitem{Aceti:2012cb}
F.~Aceti, R.~Molina, and E.~Oset.
\newblock {The $X(3872) \to J/\psi \gamma$ decay in the $D \bar D^*$ molecular
  picture}.
\newblock {\em Phys. Rev.}, D86:113007, 2012.

\bibitem{Nagahiro:2008um}
H.~Nagahiro, J.~Yamagata-Sekihara, E.~Oset, S.~Hirenzaki, and R.~Molina.
\newblock {The gamma gamma decay of the f(0)(1370) and f(2)(1270) resonances in
  the hidden gauge formalism}.
\newblock {\em Phys. Rev.}, D79:114023, 2009.

\bibitem{Ablikim:2013mio}
M.~Ablikim et~al.
\newblock {Observation of a Charged Charmoniumlike Structure in $e^+e^-$ →
  $π^+π^-$ J/ψ at $\sqrt{s}$ =4.26 GeV}.
\newblock {\em Phys. Rev. Lett.}, 110:252001, 2013.

\bibitem{Liu:2013dau}
Z.~Q. Liu et~al.
\newblock {Study of $e^+e^- → π^+ π^- J/ψ$ and Observation of a Charged
  Charmoniumlike State at Belle}.
\newblock {\em Phys. Rev. Lett.}, 110:252002, 2013.

\bibitem{Xiao:2013iha}
T.~Xiao, S.~Dobbs, A.~Tomaradze, and Kamal~K. Seth.
\newblock {Observation of the Charged Hadron $Z_c^{\pm}(3900)$ and Evidence for
  the Neutral $Z_c^0(3900)$ in $e^+e^-\to \pi\pi J/\psi$ at $\sqrt{s}=4170$
  MeV}.
\newblock {\em Phys. Lett.}, B727:366--370, 2013.

\bibitem{Ablikim:2013xfr}
M.~Ablikim et~al.
\newblock {Observation of a charged $(D\bar{D}^{*})^\pm$ mass peak in
  $e^{+}e^{-} \to \pi D\bar{D}^{*}$ at $\sqrt{s} =$ 4.26 GeV}.
\newblock {\em Phys. Rev. Lett.}, 112(2):022001, 2014.

\bibitem{Ablikim:2015swa}
M.~Ablikim et~al.
\newblock {Confirmation of a charged charmoniumlike state $Z_c(3885)^{\mp}$ in
  $e^+e^-\to\pi^{\pm}(D\bar{D}^*)^\mp$ with double $D$ tag}.
\newblock {\em Phys. Rev.}, D92(9):092006, 2015.

\bibitem{Prelovsek:2013xba}
Sasa Prelovsek and Luka Leskovec.
\newblock {Search for $Z^{+}_{c}$(3900) in the $1^{+-}$ Channel on the
  Lattice}.
\newblock {\em Phys. Lett.}, B727:172--176, 2013.

\bibitem{Prelovsek:2014swa}
Sasa Prelovsek, C.~B. Lang, Luka Leskovec, and Daniel Mohler.
\newblock {Study of the $Z_c^+$ channel using lattice QCD}.
\newblock {\em Phys. Rev.}, D91(1):014504, 2015.

\bibitem{Chen:2014afa}
Ying Chen et~al.
\newblock {Low-energy scattering of the $(D\bar{D}^*)^\pm$ system and the
  resonance-like structure $Z_c(3900)$}.
\newblock {\em Phys. Rev.}, D89(9):094506, 2014.

\bibitem{Albaladejo:2016jsg}
Miguel Albaladejo, Pedro Fernandez-Soler, and Juan Nieves.
\newblock {$Z_c(3900)$: Confronting theory and lattice simulations}.
\newblock {\em Eur. Phys. J.}, C76(10):573, 2016.

\bibitem{Abazov:2018cyu}
Victor~Mukhamedovich Abazov et~al.
\newblock {Evidence for $Z_c^{\pm}(3900)$ in semi-inclusive decays of
  $b$-flavored hadrons}.
\newblock {\em Phys. Rev.}, D98(5):052010, 2018.

\bibitem{Yuan:2018inv}
Chang-Zheng Yuan.
\newblock {The XYZ states revisited}.
\newblock {\em Int. J. Mod. Phys.}, A33(21):1830018, 2018.

\bibitem{Bando:1987br}
Masako Bando, Taichiro Kugo, and Koichi Yamawaki.
\newblock {Nonlinear Realization and Hidden Local Symmetries}.
\newblock {\em Phys. Rept.}, 164:217--314, 1988.

\bibitem{Oset:2009vf}
E.~Oset and A.~Ramos.
\newblock {Dynamically generated resonances from the vector octet-baryon octet
  interaction}.
\newblock {\em Eur. Phys. J.}, A44:445--454, 2010.

\bibitem{Aceti:2014kja}
Francesca Aceti, Melahat Bayar, Jorgivan~Morais Dias, and Eulogio Oset.
\newblock {Prediction of a $Z_c(4000)$ $D^* \bar D^*$ state and relationship to
  the claimed $Z_c(4025)$}.
\newblock {\em Eur. Phys. J.}, A50:103, 2014.

\bibitem{Meissner:1987ge}
Ulf-G. Mei{\ss}ner.
\newblock {Low-Energy Hadron Physics from Effective Chiral Lagrangians with
  Vector Mesons}.
\newblock {\em Phys. Rept.}, 161:213, 1988.

\bibitem{Passarino:1978jh}
G.~Passarino and M.~J.~G. Veltman.
\newblock {One Loop Corrections for e+ e- Annihilation Into mu+ mu- in the
  Weinberg Model}.
\newblock {\em Nucl. Phys.}, B160:151--207, 1979.

\bibitem{Aceti:2015zva}
F.~Aceti, J.~M. Dias, and E.~Oset.
\newblock {f$_{1}$(1285) decays into a$_{0}$(980)π$^{0}$, f$_{0}$(980)π$^{0}$
  and isospin breaking}.
\newblock {\em Eur. Phys. J.}, A51(4):48, 2015.

\bibitem{Wess:1971yu}
J.~Wess and B.~Zumino.
\newblock {Consequences of anomalous Ward identities}.
\newblock {\em Phys. Lett.}, 37B:95--97, 1971.

\bibitem{Witten:1983tw}
Edward Witten.
\newblock {Global Aspects of Current Algebra}.
\newblock {\em Nucl. Phys.}, B223:422--432, 1983.

\bibitem{Oh:2000qr}
Yong-seok Oh, Taesoo Song, and Su~Houng Lee.
\newblock {J / psi absorption by pi and rho mesons in meson exchange model with
  anomalous parity interactions}.
\newblock {\em Phys. Rev.}, C63:034901, 2001.

\bibitem{Nagahiro:2008mn}
H.~Nagahiro, L.~Roca, and E.~Oset.
\newblock {Meson loops in the $f_0(980)$ and $a_0(980)$ radiative decays into
  $\rho$, $\omega$}.
\newblock {\em Eur. Phys. J.}, A36:73--84, 2008.

\bibitem{Nagahiro:2008cv}
H.~Nagahiro, L.~Roca, A.~Hosaka, and E.~Oset.
\newblock {Hidden gauge formalism for the radiative decays of axial-vector
  mesons}.
\newblock {\em Phys. Rev.}, D79:014015, 2009.

\bibitem{Ozaki:2007ka}
S.~Ozaki, H.~Nagahiro, and A.~Hosaka.
\newblock {Magnetic interaction induced by the anomaly in
  kaon-photoproductions}.
\newblock {\em Phys. Lett.}, B665:178--181, 2008.

\bibitem{Torres:2014fxa}
A.~Martinez~Torres, K.~P. Khemchandani, F.~S. Navarra, M.~Nielsen, and
  Luciano~M. Abreu.
\newblock {On $X(3872)$ production in high energy heavy ion collisions}.
\newblock {\em Phys. Rev.}, D90(11):114023, 2014.
\newblock [Erratum: Phys. Rev.D93,no.5,059902(2016)].

\bibitem{Abreu:2016qci}
L.~M. Abreu, K.~P. Khemchandani, A.~Martinez~Torres, F.~S. Navarra, and
  M.~Nielsen.
\newblock {$X(3872)$ production and absorption in a hot hadron gas}.
\newblock {\em Phys. Lett.}, B761:303--309, 2016.

\bibitem{MartinezTorres:2017eio}
A.~Mart{\'\i}nez~Torres, K.~P. Khemchandani, L.~M. Abreu, F.~S. Navarra, and
  M.~Nielsen.
\newblock {Absorption and production cross sections of $K$ and $K^*$}.
\newblock {\em Phys. Rev.}, D97(5):056001, 2018.

\bibitem{Abreu:2017cof}
L.~M. Abreu, K.~P. Khemchandani, A.~Mart{\'\i}nez~Torres, F.~S. Navarra, and
  M.~Nielsen.
\newblock {Update on $J/\psi$ regeneration in a hadron gas}.
\newblock {\em Phys. Rev.}, C97(4):044902, 2018.

\end{thebibliography}

\end{document}